%% file: a.tex
\documentclass[10pt,twocolumn,letterpaper]{article}

\usepackage[accsupp]{axessibility}  % Improves PDF readability for those with disabilities.

\usepackage{cvpr}
\usepackage{float}
\usepackage{multirow,tabularx}
\usepackage{graphicx}
\usepackage{amsmath}
\usepackage{amssymb}
\usepackage{booktabs}
\usepackage{purudefs}
\usepackage{caption}
\usepackage{subcaption}
\usepackage{tikz}
\usetikzlibrary{shapes.geometric}
\usetikzlibrary{calc,shadows}
\usepackage[pagebackref,breaklinks,colorlinks]{hyperref}

\usepackage[capitalize]{cleveref}
\crefname{section}{Sec.}{Secs.}
\Crefname{section}{Section}{Sections}
\crefname{appendix}{App.}{Apps.}
\Crefname{appendix}{Appendix}{Appendices}
\crefname{table}{Tab.}{Tabs.}
\Crefname{table}{Table}{Tables}
\crefname{figure}{Fig.}{Figs.}
\Crefname{figure}{Table}{Tables}

\author{Anshul Mittal\\
Kunal Dahiya\thanks{Equal contribution. Author names appear in alphabetical order.} \\
{\small IIT Delhi} \\
{\tt\small me@anshulmittal.org} \\
{\tt\small kunalsdahiya@gmail.com}
\and
Shreya Malani\footnote[1]{} \\
Janani Ramaswamy \\
Seba Kuruvilla \\
{\small Microsoft} \\
{\tt\small \{shreya0510,seba.2312\}@gmail.com} \\
{\tt\small jramaswamy@microsoft.com}
\and
Jitendra Ajmera \\
Keng-hao Chang \\
{\small Microsoft} \\
{\tt\small jiajmera@microsoft.com} \\
{\tt\small kenchan@microsoft.com}
\and
Sumeet Agarwal \\
{\small IIT Delhi} \\
{\tt\small sumeet@iitd.ac.in}
\and
Purushottam Kar \\
{\small IIT Kanpur} \\
{\tt\small purushot@cse.iitk.ac.in}
\and
Manik Varma \\
{\small Microsoft Research, IIT Delhi} \\
{\tt\small manik@microsoft.com}
\and
\textit{In fond memory of Sh. Vinay Mittal (1963 - 2022)}
}

\newcommand{\mytitle}{Multi-modal Extreme Classification}

\newcommand{\acc}{at least 3\%\xspace}
\newcommand{\ova}{one-vs-all\xspace}
\newcommand{\Ova}{One-vs-all\xspace}

\definecolor{silver}{rgb}{0.75, 0.75, 0.75}
\newcommand{\wpred}[1]{
\textcolor{silver}{#1}\xspace
}
\newrobustcmd*{\mycircle}[1]{\tikz{\filldraw[draw=#1,fill=#1] (0,0) circle [radius=0.1cm];}}

\newrobustcmd*{\mytriangle}[1]{\tikz{\filldraw[draw=#1,fill=#1] (0,0) --(0.2cm,0) -- (0.1cm,0.2cm);}}

\newrobustcmd*{\mystar}[1]{
  \tikz{\node[draw,star,scale=0.35, star point ratio=2.25, draw=#1,fill=#1]{};}
}

\graphicspath{{figs/}}

\newcommand{\alg}{MUFIN\xspace}
\newcommand{\anns}{\text{NN}^x}
\newcommand{\code}{\href{https://github.com/Extreme-classification/MUFIN}{\color{blue}{https://github.com/Extreme-classification/MUFIN}}}\xspace

\newcommand{\suppurl}{http://manikvarma.org/pubs/mittal22-supp.pdf}

\newcommand{\repourl}{http://manikvarma.org/downloads/XC/XMLRepository.html}

\newcommand{\suppurlcite}{\url{\suppurl}}
\newcommand{\repourlcite}{\url{\repourl}}
\newcommand{\suppcite}{supplementary \href{\suppurl}{[link]}\xspace}
\newcommand{\repocite}{The Extreme Classification Repository \cite{XMLRepo} \href{\repourl}{[link]}\xspace}
\newcommand{\suppciteshort}{\href{\suppurl}{supplementary}\xspace}

\def\cvprPaperID{1836}
\def\confName{CVPR}
\def\confYear{2022}

\begin{document}

\title{\mytitle}

\maketitle

\input{abstract}

\newpage

\input{intro}
\input{related}
\input{method}
\input{exps}
\input{supp}

\section*{Acknowledgements}
The authors thank the reviewers for helpful comments. The authors are grateful to the High-Performance Computing facility and staff at IIT Delhi. AM is supported by a Google PhD Fellowship. AM, KD, and SA acknowledge funding from a Microsoft Research India grant on Extreme Classification. PK is supported by Microsoft Research India via consultancy grant no MRLIPL/CS/2021296.

\clearpage
{\small
\bibliographystyle{ieee_fullname}
\bibliography{refs}
}
% \clearpage
\input{app}
\end{document}

%% file: abstract.tex
\begin{abstract}
    This paper develops the \alg technique for extreme classification (XC) tasks with millions of labels where datapoints and labels are endowed with visual and textual descriptors. Applications of \alg to product-to-product recommendation and bid query prediction over several millions of products are presented. Contemporary multi-modal methods frequently rely on purely embedding-based methods. On the other hand, XC methods utilize classifier architectures to offer superior accuracies than embedding-only methods but mostly focus on text-based categorization tasks. \alg bridges this gap by reformulating multi-modal categorization as an XC problem with several millions of labels. This presents the twin challenges of developing multi-modal architectures that can offer embeddings sufficiently expressive to allow accurate categorization over millions of labels; and training and inference routines that scale logarithmically in the number of labels. \alg develops an architecture based on cross-modal attention and trains it in a modular fashion using pre-training and positive and negative mining. A novel product-to-product recommendation dataset MM-AmazonTitles-300K containing over 300K products was curated from publicly available amazon.com listings with each product endowed with a title and multiple images. On the all datasets \alg offered \acc higher accuracy than leading text-based, image-based and multi-modal techniques. Code for \alg is available at \code
\end{abstract}

%% file: intro.tex
\begin{figure}[H]
    \centering
    \includegraphics[width=\columnwidth]{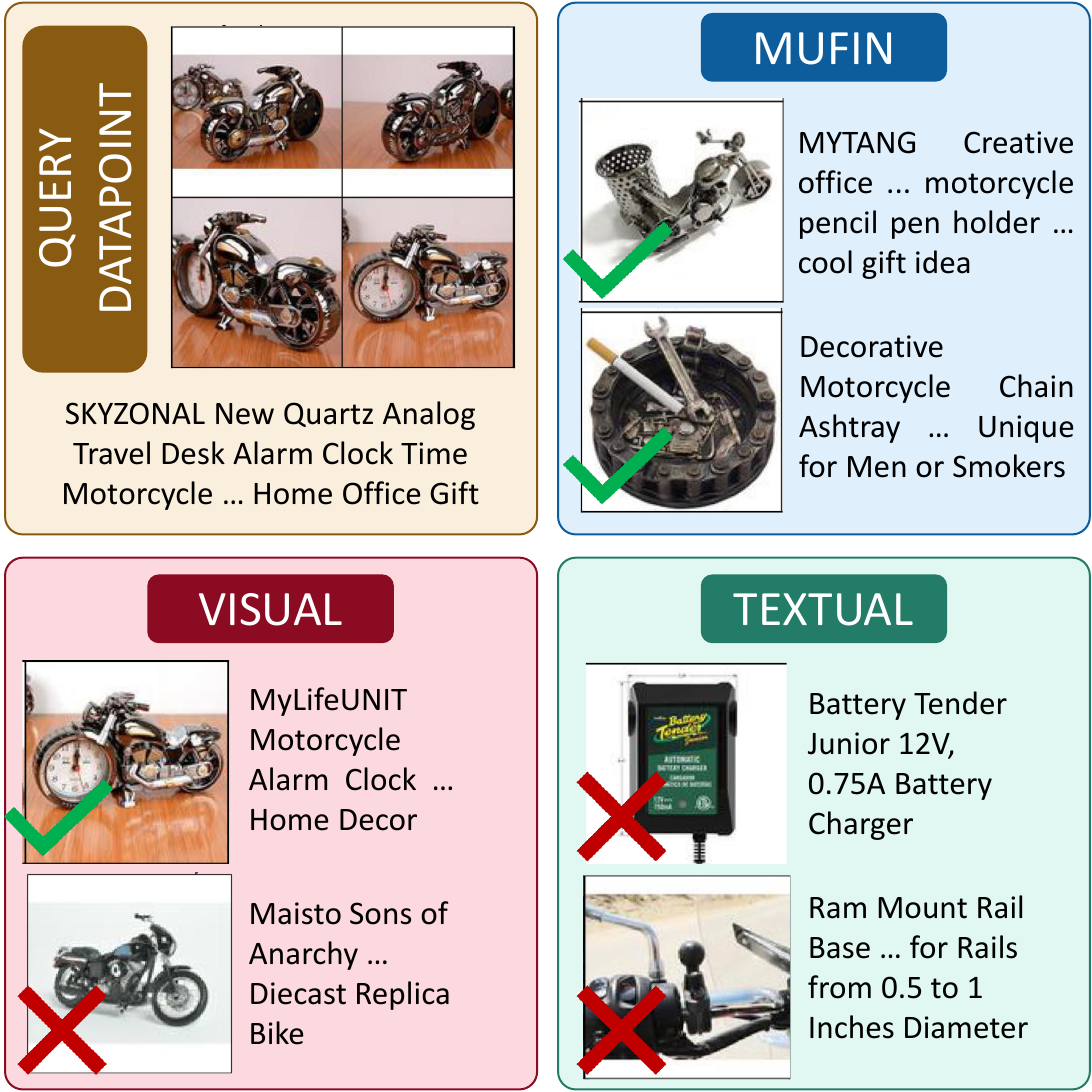}
    \caption{Predictions on the MM-AmazonTitles-300K product-to-product recommendation task illustrate the need for accurate multi-modal retrieval. For a decorative motorcycle-shaped alarm clock as the query product, multi-modal retrieval using \alg was able to retrieve visually similar products such as a motorcycle-shaped pencil holder as well as visually dissimilar but related products such as a motorcycle themed ashtray. Recovery using the visual modality alone ignored thematically linked products, instead recovering mostly motorcycle-shaped products. Textual recovery on the other hand fixated on the word ``motorcycle" and started recovering accessories for actual motorcycles.}
    \label{fig:teaser}
\end{figure}

\section{Introduction}
\label{sec:intro}

\noindent\textbf{Extreme Classification (XC).} The goal of extreme multi-label classification is to develop architectures to annotate datapoints with the most relevant \textit{subset} of labels from an extremely large set of labels. For instance, given a product purchased by a user, we may wish to recommend to the user, the subset (i.e. one or more) of the most related products from an extremely large inventory of products. In this example, the purchased product is the datapoint and each product in the inventory becomes a potential label for that datapoint. Note that multi-label classification generalizes multi-class classification where the objective is to predict a single mutually exclusive label for a given datapoint. An example of a multi-class problem would be to assign a product to a single exclusive category in a product taxonomy.

% \noindent\textbf{Multi-modal Data.} An interesting XC application arises when datapoints and labels are endowed with multiple descriptors that can be visual and textual. A prominent real-life use case is product-to-product recommendation \cite{Mittal21b} where product titles and product details are available as textual descriptors and one or more product images are available as visual descriptors. Another use-case is bid-query prediction~\cite{Dahiya21b} where the goal is to take an advertisement with visual and textual descriptions and tag it with the list of (textual) user queries that are most likely to lead to a click on that ad. A third use-case is that of identifying compatible outfits where each outfit is described using multiple images and a textual caption \cite{Vasileva2018}.

\noindent\textbf{Multi-modal XC.} An interesting XC application arises when datapoints and labels are endowed with both visual and textual descriptors. Example uses cases include\\
(1) Product-to-product recommendation \cite{Mittal21b} with products being represented using their titles and one or more images.\\
(2) Bid-query prediction~\cite{Dahiya21b} where an advertisement with visual and textual descriptions has to be tagged with the list of user queries most likely to lead to a click on that ad.\\
(3) Identifying compatible outfits where each outfit is described using multiple images and a textual caption \cite{Vasileva2018}.

\noindent\textbf{Challenges in Multi-modal XC.} Existing multi-modal methods~\cite{Hou2021,Revanur2021,Tan2019,Vasileva2018} are often \emph{embeddings-only} i.e. categorization is done entirely using embeddings of datapoints and categories obtained from some neural architecture. However, XC research has shown that training classifiers alongside embedding architectures can offer improved results~\cite{Chang20,Dahiya21b,You18}. However, existing XC research focuses mostly on text-based categorization. Bridging this gap requires architectures that offer multi-modal embeddings sufficiently expressive to perform categorization over millions of classes. Also required are routines that can train classifiers over millions of classes and still offer predictions in milliseconds as demanded by real-time applications \cite{Chang20,Dahiya21b,Jain19}. This is usually possible only if training and inference scale logarithmically with the number of labels.

% \noindent\textbf{Contributions.} This paper presents the \alg technique for XC tasks where both datapoints and labels can be endowed with multiple descriptors -- both visual and textual.\\
% 1. \alg introduces a novel architecture that melds multi-modal attention into a scalable extreme classification framework with high-capacity \ova classifiers and datapoint-label cross attention.\\
% 2. \alg' proposed training framework scales to tasks with millions of labels by effectively using pre-training and hard-positive and hard-negative mining techniques.\\
% 3. \alg could offer predictions within 3-4 milliseconds per test point even on tasks with millions of labels.\\
% 4. \alg's multi-modal cross-attention framework seamlessly extends to tasks with unseen labels, offering superior performance than state-of-the-art techniques.\\
% 5. \alg's architecture does not insist on a fixed number of descriptors and is resilient to the visual or textual mode going entirely missing for a datapoint or label.\\
% 6. To accurately benchmark multi-modal XC methods, the MM-AmazonTitles-300K product-to-product recommendation dataset is curated (and released with this submission) from publicly available amazon.com listings with over 300K products each having a title and multiple images.\\
% 7. On multiple datasets including Polyvore-Disjoint, MM-AmazonTitles-300K, and a dataset A2Q-4M with more than 4M labels curated from click logs of a popular search engine, MUFIN could offer accuracies \acc higher than leading text-, image- and multi-modal-based techniques.

\noindent\textbf{Contributions.} The \alg method targets XC tasks with millions of labels where both datapoints and labels can be endowed with visual and textual descriptors.\\
(1) \alg melds a novel embedding architecture and a novel classifier architecture. The former uses multi-modal attention whereas the latter uses datapoint-label cross attention and high-capacity \ova classifiers.\\
(2) \alg training scales to tasks with several millions of labels by using pre-training and hard-positive and hard-negative mining. \alg offers predictions within 3-4 milliseconds per test point even on tasks with millions of labels.\\
(3) This paper releases the MM-AmazonTitles-300K product-to-product recommendation dataset curated from publicly available amazon.com listings with over 300K products each having a title and multiple images.\\
(4) \alg offers \acc higher accuracy than leading text-only, image-only and multi-modal methods on several tasks (MM-AmazonTitles-300K, A2Q-4M) including zero-shot tasks (Polyvore) indicating the superiority of not just \alg's classifiers but its embedding model as well.

% (5) A diverse range of ablation studies confirm that \alg's design choices with respect to architecture design and training pipeline are indeed optimal.

%% file: related.tex
\section{Related Work}
\label{sec:related}
% Below we survey related work along four directions of immediate interest to this paper.

\noindent\textbf{Large-scale Visual Categorization.} Categorization with a large number of classes has received much attention~\cite{liu2021a, Huynh2021,Guo_2021_CVPR}. Early methods learnt classifiers over hand-crafted or pre-trained features such as HoG \cite{Deanetal2013} ($100K$ classes). Contemporary approaches offer superior accuracies by using task-specific representations obtained from neural architectures. Some of these \cite{radford2021,Wu17,Guo_2021_CVPR,liu2021a} eschew classifiers entirely and focus on purely embedding-based methods while others train embedding and classifier models jointly using techniques such as hierarchical soft-max, in-batch negative mining \cite{Joulin2016} and hard-negative mining \cite{Zhuetal19}. However, these works do not consider multi-modal data.

% Approaches to scale  to tasks with several tens of thousands of categories or more e.g. ImageNet ($> 20K$ categories) \cite{Deng2009}, Yolo9000 ($> 9K$ categories) \cite{Redmon16}, YFCC100M ($> 100K$ categories) \cite{Thomee16,Joulin2016} and CASIA ($> 2M$ categories) \cite{Zhuetal19}  Contemporary approaches utilizing representation learning utilize techniques such as  to reduce training time.  However, these works mostly do not consider multi-modal data.

\noindent\textbf{Extreme Classification.} XC methods seek to learn classifiers that offer efficient prediction even with millions of labels. Earlier works used fixed or pre-trained features and learnt classifier architectures such as multi-way classification trees \cite{Khandagale19}, \ova classifiers \cite{Jain19,Babbar17} and probabilistic label trees \cite{Jasinska16}. Recent advances \cite{Dahiya23b, Dahiya23, Dahiya21, Dahiya21b,Mittal21b,You18,Kharbanda21,Jiang21,Saini21} have introduced task-specific neural representations that are jointly learnt alongside the classifiers and offer performance boosts over embedding-only methods. However, these mostly consider tasks with textual descriptions only.

\noindent\textbf{Multi-modal Product Recommendation.} The task of recommending related products such as compatible outfits \cite{Vasileva2018} has led to several multi-modal techniques that utilize product images as well as product title or category. ADDE-O \cite{Hou2021} learns a disentangled visual representation for outfits so that an outfit with an altered category such as color or size can be recovered simply by appending the query product with a category modifier such as ``blue'' or ``extra large''. The Type-aware approach \cite{Vasileva2018} learns product embeddings that respect textual product types but capture product similarity and compatibility. SCE-Net \cite{Tan2019} learns image representations that jointly capture multiple aspects of similarity e.g. color, texture without having to learn separate feature spaces for each aspect. SSVR \cite{Revanur2021} introduces semi- and self-supervised techniques that use textual categories to regularize product image embeddings. S-VAL \cite{Kim2021} and CSA-Net \cite{lin2020} perform similarity and compatibility-based retrieval focusing on using the visual modality alone or else using the textual category/type information as a black-box category. Note that none of these methods utilize classifiers and are purely embedding-based methods. Modality fusion techniques have also been explored. Early works adopted late fusion by treating modalities separately till each yielded a score whereas recent works \cite{Nagrani21} have explored early and \textit{bottle-necked} fusion. \alg performs early fusion via its multi-modal attention blocks (see \cref{sec:method}).

\noindent\textbf{Multi-Modal Learning.} Methods for multi-modal tasks such as image captioning and associated word prediction \cite{Joulin2016} have proposed embedding-only solutions (CLIP \cite{radford2021}, VisualBERT \cite{li2019}) as well as classifier architectures (IMRAM \cite{Chen20}, M3TR \cite{Zhao21}). \alg empirically outperforms CLIP and VisualBERT while IMRAM and M3TR could not scale to the datasets used in our experiments.

% \subsection{\alg's Contributions in Context.}
% \noindent\textbf{Multi-modal Classification.} \alg melds XC techniques with multi-modal embedding architectures in a manner that outperforms state-of-the-art extreme classification methods that consider only the textual modality, leading methods that consider only the visual modality, as well as leading techniques for multi- and cross-modal tasks.

% \noindent\textbf{Multi-modal Embedding.}  Experiments on datasets such as Polyvore which presents only unseen labels at test time establish that \alg's superior performance is not just a result of its high-capacity \ova classifier architecture, but also its multi-modal embedding architecture.

% \noindent\textbf{Scalability.} \alg scales to large datasets such as MM-Amazon-300K and A2Q-4M with more than 4M labels while offering prediction times within 3-4 milliseconds.

% \noindent\textbf{Ablations.} A diverse range of ablation studies confirm that \alg's design choices with respect to architecture design and training pipeline are indeed optimal.

%% file: method.tex
\section{\alg MUltimodal extreme classiFIcatioN}
\label{sec:method}

\begin{figure*}
    \centering
    \includegraphics[width=\textwidth]{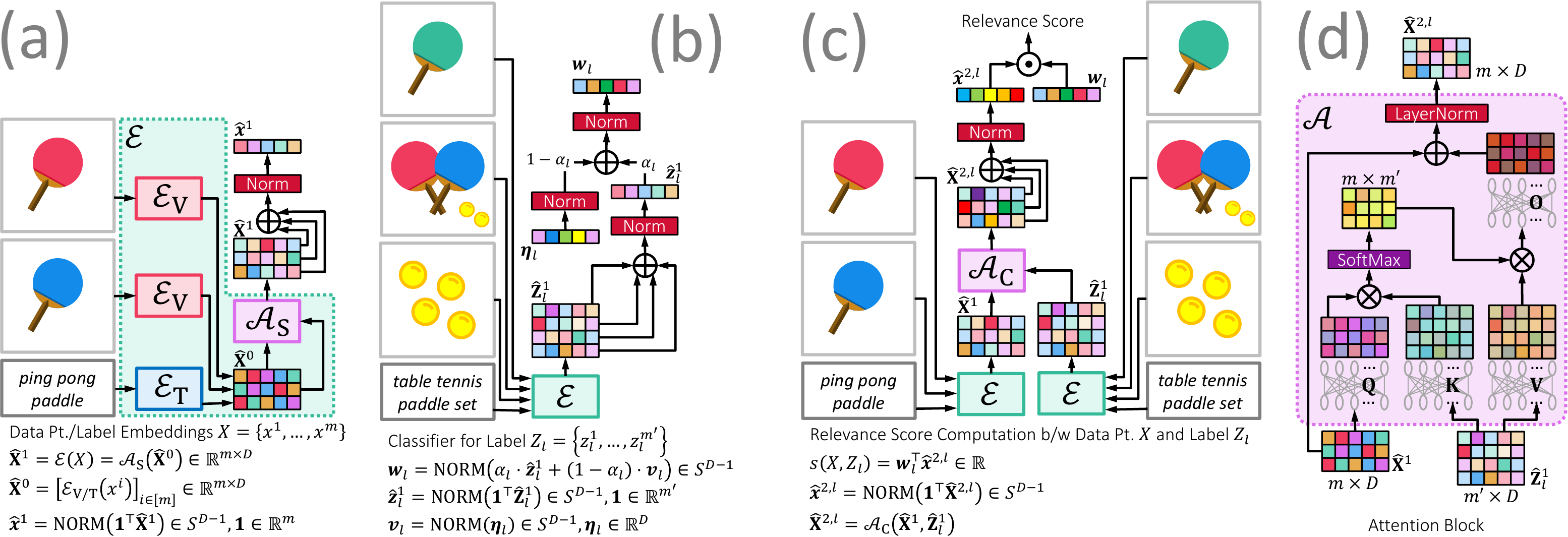}
    \caption{\textbf{(a)} The embedding block $\cE$ uses a multi-modal self-attention block $\cA_S$ to represent a datapoint (or a label) as a bag of embeddings $\hat\vX^1$ that can be optionally aggregated into a single normalized vector $\hat\vx^1$. \textbf{(b)} \Ova classifier vectors $\vw_l$ are learnt for each label by combining the vector representation for that label $\hat\vz^1_l$ with a normalized free vector $\kN(\veta_l)$ ($\kN$ is normalization). \textbf{(c)} The relevance score between a datapoint $i$ and label $l$ is computed using the label classifier $\vw_l$ and a vector representation of the datapoint $\hat\vx_i^{2,l}$ that is adapted to the label $l$ by using the cross-attention block $\cA_C$. \textbf{(d)} The attention block is instantiated twice, once to implement self-attention as $\cA_{\text S}: X \mapsto \cA(X,X)$ and once to implement cross-attention as $\cA_{\text C}: (X,Z) \mapsto \cA(X,Z)$. The two instantiations use distinct parameters.}
    \label{fig:arch}
\end{figure*}

% \begin{figure}
%     \centering
%     \includegraphics[width=0.5\columnwidth]{figs/attn-crop.pdf}
%     \caption{The attention block is instantiated by \alg twice, once to implement self-attention as $\cA_{\text S}: X \mapsto \cA(X,X)$ and once to implement cross-attention as $\cA_{\text C}: (X,Z) \mapsto \cA(X,Z)$. The two instantiations use distinct parameters.}
%     \label{fig:attn}
% \end{figure}

\noindent\textbf{Notation.} $L$ is the number of labels (\eg number of products available for recommendation, bid queries). $N$ train points are presented as $\bc{(X_i,\vy_i)}_{i=1}^N$. Datapoint $i$ is represented using $m_i$ descriptors (textual e.g. product title and/or visual e.g. product image) as $X_i = \bc{x_i^1,\ldots,x_i^{m_i}}$. $\vy_i \in \bc{-1,+1}^L$ is the ground-truth label vector for datapoint $i$ with $y_{il} = +1$ if label $l \in [L]$ is relevant to the datapoint $i$ else $y_{il} = -1$. Each label $l \in [L]$ is represented as $Z_l = \bc{z_l^1,\ldots,z_l^{m_l}}$ using $m_l$ textual/visual descriptors.

\textbf{Motivation for \alg's Architecture.} \alg seeks to obtain an embedding $\hat\vx_i \in \bR^D$ for every datapoint $X_i$ and a classifier $\vw_l \in \bR^D$ for every label $l \in [L]$ so that $\vw_l^\top\hat\vx_i$ is indicative of the relevance of label $l$ to datapoint $i$. Datapoints and labels each having multiple descriptors i.e. $m_i,m_l \geq 1$ present opportunities to ease this process:\\
(1) The neural architecture used to obtain datapoint embeddings $\hat\vx_i$ can also be used to obtain label embeddings $\hat\vz_l$ that can serve as a convenient warm start when learning $\vw_l$ and has been found to accelerate training in XC methods \cite{Mittal21b,Dahiya21b}.\\
(2) Cross-talk among descriptors of a datapoint and those of a label may make the classifier's job easier by promoting affinity among related datapoint-label pairs. Alignment between descriptors of datapoint $i$ and those of label $l$ can be used to construct an alternate embedding $\hat\vx^l_i$ of the datapoint that is \emph{adapted} to the label $l$. The goal of this \emph{label-adapted} embedding would be not budge if the label $l$ is irrelevant i.e. $\hat\vx^l_i \approx \hat\vx_i$ if $y_{il} = -1$ but approach the label classifier if the label is relevant i.e. $\hat\vx^l_i \rightarrow \hat\vw_l$ if $y_{il} = +1$.\\
(3) Self-talk among descriptors of the same datapoint/label allow different modalities to interact and produce superior embeddings for that datapoint/label.\\
\alg adopts both bag and vector representations for labels and datapoints to let descriptors retain their identity and allow efficient classification. Attention blocks are used to implement cross-talk and self-talk. \cref{fig:xattn_visuals} shows that label-adapted embeddings learnt by \alg do achieve the objectives stated above by noticing that images of a datapoint appear among images of a relevant label.
% Ablation experiments in \cref{sec:exps} show that each component in \alg's architecture makes non-trivial contribution to its overall performance.
% The architecture is detailed below and pictorially depicted in \cref{fig:arch}.

\noindent\textbf{Bag Embeddings.} A visual architecture $\cE_V$ is used to map visual descriptors to $\bR^D$ (\alg uses ViT-32~\cite{dosovitskiy2021} with $D = 192$). A textual architecture $\cE_T$ (\alg uses msmarco-distilbert-base-v4~\cite{reimers2019} with $D = 192$) is used to map textual descriptors to $\bR^D$. We note that both the ViT and Sentence-BERT models have a native dimensionality of 768. An adaptive maxpool 1D layer was use to project down to obtain 192 dimensional descriptor embeddings. $\cE_V, \cE_T$ are shared by datapoints and labels. \alg maps datapoints and labels to bags of embeddings as shown in \cref{fig:arch}(a). A datapoint $X_i = \bc{x_i^1,\ldots,x_i^{m_i}}$ is mapped to $\hat\vX^1_i = \cE(X_i) \in \bR^{m_i \times D}$ by first encoding each descriptor of that datapoint using either $\cE_V$ or $\cE_T$, depending on whether that descriptor is visual or textual, to obtain a bag of pre-embeddings $\hat\vX^0_i \in \bR^{m_i \times D}$. These are then passed through a self-attention block $\cA_S$ (an instantiation of the block depicted in \cref{fig:arch}(d)) to obtain $\hat\vX^1_i = \cA_S(\hat\vX^0_i) = \cA(\hat\vX^0_i,\hat\vX^0_i)$. A label $Z_l = \bc{z_l^1,\ldots,z_l^{m_l}}$ is similarly mapped to $\hat\vZ^1_l = \cE(Z_l) \in \bR^{m_l \times D}$. We note that the same self-attention block $\cA_S$ is used to embed both datapoints and labels.

\noindent\textbf{Vector Embeddings.} \alg obtains vector embeddings by aggregating and normalizing the bag embeddings offered by $\cE$ (see \cref{fig:arch}(a)). The vector embedding for a datapoint $i$ is obtained as $\hat\vx_i^1 = \kN(\vone^\top\hat\vX_i^1) \in S^{D-1}$ where $\vone \in \bR^{m_i}$ is the all ones vector, $\kN: \vv \mapsto \vv/\norm\vv_2$ is the normalization operator and $S^{D-1}$ is the unit sphere in $D$ dimensions. Similarly $\hat\vz_l^1 = \kN(\vone^\top\hat\vZ_l^1)$. Given a datapoint $i$ and label $l$, \alg constructs the label-adapted embedding $\hat\vx^{2,l}_i$ for the datapoint as shown in \cref{fig:arch}(c). A bag embedding for the datapoint adapted to the label is obtained as $\hat\vX_i^{2,l} = \cA_C(\hat\vX^1_i, \hat\vZ^1_l)$ where $\hat\vX^1_i = \cE(X_i), \hat\vZ^1_l = \cE(Z_l)$ using a cross-attention block $\cA_C$ (\cref{fig:arch}(d)) which is vectorized to yield the label-adapted vector embedding $\hat\vx^{2,l}_i = \kN(\vone^\top\hat\vX^{2,l}_i)$. Note that $\cA_S$ and $\cA_C$ do not share parameters.

\noindent\textbf{Scoring Model and Label Classifiers.} Given a datapoint $i$ and a label $l \in [L]$, \alg assigns a relevance score by taking a dot product of the adapted vector embedding of the datapoint $\hat\vx_i^{2,l}$ with the classifier vector $\vw_l$ constructed as shown in \cref{fig:arch}(b) by linearly combining the (normalized) vector embedding for the label $\hat\vz^1_l$ with a normalized free vector $\kN(\veta_l)$. The free vector $\veta_l$ and the combination weight $\alpha_l \in [0,1]$ are learnt independently per label.
% For a label $l \notin [L]$ not seen during training, we simply set $\alpha_l = 1$ to let the label embedding itself act as the classifier.

\subsection{Modular Training with \alg}

\noindent\textbf{Trainable Parameters.} The encoder blocks $\cE_V, \cE_T$, the attention blocks $\cA_S, \cA_C$, the free vectors and weights $\veta_l, \alpha_l, l \in [L]$ for the label classifiers were trained. \alg adopted a training strategy first proposed in the DeepXML paper~\cite{Dahiya21} that breaks training into 4 distinct modules.

\noindent\textbf{Module I: Pre-training.} In this module, only the encoders $\cE_V, \cE_T$ and the self-attention block $\cA_S$ were trained in a Siamese fashion. The cross-attention block $\cA_C$ was bypassed i.e $\hat\vx^{2,l}_i = \hat\vx^1_i$ and $\alpha_l$ was set to $0$ for all $l \in [L]$ so as to also exclude the free vectors. A pretrained ViT-32 model \cite{dosovitskiy2021} was used to initialize $\cE_V$ and its final layer was fine-tuned during training. A pre-trained SentenceBERT model (msmarco-distilbert-base-v4) \cite{reimers2019} was used to initialize $\cE_T$ and was fine-tuned end-to-end during training. The transformation layers $Q, K, V, O$ in $\cA_S$ were initialized to identity. Datapoints and labels were represented by their vector embeddings i.e. $\hat\vx^1_i$ and $\hat\vz^1_l$ respectively. Training encouraged $\hat\vx^1_i$ and $\hat\vz^1_l$ to approach each other for related pairs and repel for unrelated pairs. Mini-batches $B$ were created over labels instead of datapoints by sampling labels randomly. This was observed to improve performance over rare labels~\cite{Dahiya21b,Mikolov13}. Training with respect to all $N$ datapoints for each label would have resulted in an $\Om{NL}$ epoch complexity that is infeasible when $N,L$ are both in the millions. Thus, a set $\cP_l$ of \textit{hard-positive} datapoints for each label $l \in B$ was chosen among the set $\bc{i: y_{il} = +1, \ip{\hat\vz^1_l}{\hat\vx^1_i} \leq 0.9}$ since positive datapoints too similar to the label i.e. $\ip{\hat\vz^1_l}{\hat\vx^1_i} > 0.9, y_{il} = +1$ would yield vanishing gradients. In-batch negative sampling was also done by selecting for each label $l \in B$, a set $\cN_l$ of \textit{hard-negative} datapoints among positive datapoints of other labels in the same minibatch. Hard-positive and negative mining was found to accelerate training by focusing on those label-datapoint pairs that gave most prominent gradients. The following contrastive loss was used to train $\cE_V, \cE_T$ and $\cA_S$ using mini-batches $B$ over labels:
\[
\sum_{l \in B}\sum_{i \in \cP_l}\sum_{j \in \cN_l}\bs{\ip{\hat\vz^1_l}{\hat\vx^1_j} - \ip{\hat\vz^1_l}{\hat\vx^1_i} + \gamma}_+
\]
\alg used $\gamma = 0.2, \abs{\cP_l} = 2, \abs{\cN_l} = 3$. Capping the sizes of the sets to $\abs{\cP_l},\abs{\cN_l} \leq \bigO1$ ensured that an epoch complexity of $\bigO L$ instead of $\Om{LN}$.

% Training with respect to all positive and negative datapoints of a label would have resulted in an epoch requiring $\Om{NL}$ time that is infeasible when $N,L$ are both in the millions. To remedy this, for each label $l \in B$, a set $\cN_l$ of \textit{hard-negative} datapoints was chosen by executing an ANNS (approximate nearest neighbor search) query over an HNSW \cite{MalkovY16} structure constructed over datapoint embeddings $\bc{\hat\vx_i^1: i \in [N]}$. The ANNS structure was refreshed regularly (every 5 epochs) as suggested by \cite{Xiong20}.

\noindent\textbf{Module II: Augmented Retrieval.} The label-wise training strategy adopted by Module I is sympathetic to rare labels but is not aligned to final prediction where labels need to be predicted for datapoints, not the other way round. Moreover, in-batch negative mining is inexpensive but may offer inferior convergence \cite{Xiong20}. To accelerate subsequent training, a set of $\bigO{\log L}$ most promising labels was retrieved for each datapoint. The irrelevant labels in this set would form hard-negatives for subsequent training. \alg improved retrieval by exploiting multiple descriptors for each label. After Module I, datapoint vector and label bag embeddings i.e. $\hat\vx^1_i, \hat\vZ^1_l$ were re-computed. Label centroid vectors~\cite{Dahiya21b} were created as $\hat\vmu_l = \text{mean}\bc{\hat\vx^1_i: y_{il} = +1}$. An ANNS (approximate nearest neighbor search) structure $\anns$ \cite{MalkovY16} was created over the set of $\sum_{l \in [L]} (m_l+1)$ vectors $\bigcup_{l\in[L]}\hat\vZ^1_l \cup \bc{\hat\vmu_l}$ with each vector recording the identity of the label to which it belonged. ANNS queries of the form $\anns(\hat\vx^1_i)$ were then fired to retrieve for each datapoint $i$, a set $R_i$ of $\bigO{\log L} \leq 100$ unique labels. Negative labels in this set i.e. $\bc{l \in R_i: y_{il} = -1}$ were well-suited to serve as hard-negative labels for the datapoint $i$. Ablations in \cref{sec:exps} show that this technique offers superior performance than if the ANNS structure $\anns$ were to be created over vector representations $\hat\vz^1_l$ of the labels instead. Note that we could have fired bag-queries on the datapoint side as well i.e. fire $m_i$ ANNS queries for datapoint $i$, one for each element of the datapoint bag $\hat\vX^1_i = \cE(X_i)$. However that would substantially increase retrieval time by a factor of $m_i$ ($m_i \approx 5$ for MM-Amazon-300K) and was thus avoided. Instead, the approach adopted by \alg ensures superior retrieval at the cost of a single ANNS query per datapoint.

\noindent\textbf{Module III: Pre-training to Fine-tuning Transfer.} The encoders $\cE_V,\cE_T$ and the self-attention block $\cA_S$ were initialized to their values after Module I training. The transformation layers $Q, K, V, O$ within $\cA_C$ are initialized to identity. The free vectors $\veta_l$ were all offered uniform Xavier initialization and $\alpha_l = 0.5$ was initialized for all labels $l \in [L]$.

\noindent\textbf{Module IV: Fine-tuning.} $\cE_V,\cE_T,\cA_S$ were further fine-tuned whereas $\cA_C, \veta_l, l \in [L]$ and $\alpha_l, l \in [L]$ were trained from scratch. Mini-batches $B$ were created over datapoints to align with the final prediction task. For each $i \in B$, a set $\cS_i$ of random positive labels was chosen. A set $\cT_i$ of hard-negative labels was chosen among negative labels in the shortlist $R_i$ constructed in Module II. A datapoint $i$ was represented using label-adapted embeddings w.r.t. the positive and hard-negative labels shortlisted for them i.e. $\bc{\hat\vx_i^{2,l}: l \in \cS_i \cup \cT_i}$. Labels were represented using their classifier vectors $\vw_l$ (see \cref{fig:arch}(b)). The following cosine embedding loss was used to train $\cE_V,\cE_T,\cA_S,\cA_C$ and $\veta_l, \alpha_l, l \in [L]$ using mini-batches $B$ of datapoints:
\begin{align*}
    \sum_{i \in B}\bc{\sum_{l \in \cS_i} \br{1 - \ip{\hat\vx_i^{2,l}}{\vw_l}} + \sum_{k \in \cT_i}\bs{\ip{\hat\vx_i^{2,k}}{\vw_k} - \gamma}_+}
\end{align*}
\alg used $\gamma = 0.5, \abs{\cS_i} = 2, \abs{\cT_i} = 12$. Capping the set sizes to $\abs{\cS_i},\abs{\cT_i} \leq \bigO{\log L}$ ensured that an epoch complexity of $\bigO{N\log L}$ instead of $\Om{NL}$.

\noindent\textbf{Prediction with \alg.} Given a test point $X_t$ with $m_t$ descriptors $X_t = \bc{x^1_t,\ldots,x^{m_t}_t}$, its vector representation $\hat\vx^1_t = \kN\br{\vone^\top\cE(X_t)}$ is used to query the ANNS structure and perform augmented retrieval of labels to yield a shortlist $R_t=\anns(\hat\vx^1_t)$ of $100 \leq \bigO{\log L}$ labels. For each retrieved label $l \in R_t$, a \emph{similarity} score is assigned as $a_{tl} \deff \max\ \ip{\hat\vx^1_t}{\vv}, \vv \in \hat\vZ^1_l \cup \bc{\hat\vmu_l}$ (recall that in augmented retrieval, each label $l \in [L]$ contributes $m_l + 1$ entries to the ANNS structure). Vector representations for $X_t$ adapted to all shortlisted labels i.e. $\bc{\hat\vx^{2,l}_t, l \in R_t}$ are computed and the corresponding label classifiers applied to yield \emph{classifier} scores $c_{il} \deff \ip{\vw_l}{\hat\vx^{2,l}_t}$ for each $l \in R_t$. The classifier and similarity scores are then combined linearly as $s_{tl} = \beta\cdot c_{tl} + (1-\beta)\cdot a_{tl}$. A fixed value of $\beta = 0.7$ was used. Final predictions are made in descending order of the scores $s_{tl}$. The prediction time complexity of \alg is derived in \cref{supp:pseudocode} in the \suppciteshort.

% \subsection{\alg Variants}

% \noindent\textbf{\alg-lite}: This variant offers reduced training time by freezing the encoders $\cE_V,\cE_T$ after Module I but keeping the rest of the training pipeline as described above. \alg-lite incurred a marginal reduction in performance (see \cref{tab:mm_amz_300k}) but offered $4.8\times$ reduction in training time.

\noindent\textbf{Handling unseen labels with \alg ($\valpha\boldsymbol=\vone$).} A variant dubbed ``\alg ($\alpha=1$)'' was developed to handle unseen labels (for which supervision was not available during training) by setting $\alpha_l = 1$ for all $l \in [L]$. This causes \alg to start using the vector label representation itself as the classifier i.e. $\vw_l \equiv \hat\vz^1_l$ and give relevance scores of the form $\ip{\hat\vz^1_l}{\hat\vx^{2,l}_t}$. Note that the cross-attention block $\cA_C$ can still be applied to yield adapted datapoint embeddings $\hat\vx^{2,l}_t$ even w.r.t. unseen labels. The variant \alg ($\alpha=1$) sets $\alpha_l = 1$ for all labels $l \in [L]$ to ensure consistency.

%% file: exps.tex
\section{Experimental Results}
\label{sec:exps}
% \alg's performance is reported on publicly available benchmark datasets (Polyvore-outfits and MM-AmazonTitles-300K) as well as a proprietary Ads-to-Query (A2Q-4M) dataset involving up to 4 million queries. 
\label{main:dataset}
\noindent\textbf{Datasets.} Dataset construction details are given in \cref{supp:datasets} in the \suppcite. \cref{tab:stats} presents dataset statistics. The Polyvore FITB task relies on precomputed shortlists for each query and does not present a satisfactory benchmark for multi-modal XC methods where the goal is to retrieve results directly from a catalog of millions of labels. Other multimodal datasets~\cite{lin2014, liu2016} were found similarly lacking. Thus, two other tasks were considered. The A2Q-4M dataset presents a heterogeneous task where datapoints have multi-modal descriptors but labels are purely textual. The MM-AmazonTitles-300K dataset also presents occasional datapoints/labels with either the text or vision modality missing entirely. Thus, these tasks demand that the architecture be resilient to missing modes.

% \alg could be up to 3--4\% more accurate than existing methods on this dataset.
% Each composition is rich in multi-modal information such as images, text descriptions, \etc.

\noindent\underline {\textit{MM-AmazonTitles-300K}}: An XC product-to-product recommendation dataset was curated from an Amazon click dump~\cite{ni2019}. Given a query product, the task is to retrieve the subset of the most relevant products from a catalog of over 300K unique products. Each product is represented by a title and up to 15 images. This dataset has been released at the \repocite.

% For e.g. for datapoints ``\textit{Persol Sunglasses}'' frequently bought items are ``\textit{Plastic Watch Crystal ...}'', ``\textit{Persol PO 3019s 52 MM Black Frame/crystal green Lens Sunglasses}'' \etc. Here, 

% \alg' ability to scale to millions of labels is demonstrated using the internal Ads-to-Query datasets with 4 million classes.

\noindent\underline {\textit{A2Q-4M}}: A large bid-query prediction task was mined from the internal click logs of the Bing search engine. Given an ad as a datapoint represented by an image and textual description, the task is to predict the subset of user queries (textual) most likely to lead to a click on that ad.

\noindent\underline {\textit{Polyvore-Disjoint}}: Polyvore is a popular fashion website where users can create outfit compositions~\cite{Vasileva2018}. The Fill-In-The-Blank (FITB) task requires the most compatible outfit to be chosen from a pre-computed shortlist given an incomplete \emph{query} outfit with 4--5 images and short captions.

% Only those ad-query pairs were included in the dataset that were clicked above a certain number of times.

\input{tables/dataset_stats}
\label{main:baselines}
\noindent\textbf{Baselines.} Due to lack of space, a detailed discussion on the baselines is provided in \cref{supp:baseline} of the \suppciteshort.

\noindent\underline {\textit{MM-AmazonTitles-300K}}: \alg was compared to leading text-based XC methods~\cite{You18, Mittal21b, Dahiya21b, Medini2019, Khandagale19, Wydmuch18} and leading multi-modal methods CLIP~\cite{radford2021} and VisualBert~\cite{li2019} that employ cross-modal pre-training to embed related items (\eg an image and its associated caption) closeby. AttentionXML~\cite{You18} employs label-specific datapoint representations similar to \alg. SiameseXML was augmented to utilize a DistilBERT architecture similar to \alg instead of the bag-of-embeddings model used in \cite{Dahiya21b}. The details of the augmentation are given in \cref{supp:baseline}. CLIP and VisualBert use the ViT and Resnet-101 image encoders respectively. To offer a fair comparison, pre-trained encoders for these methods were injected into \alg's training pipeline and afforded the same self-attention, cross attention and classifier architectures. \cref{tab:mm_amz_300k} shows that \alg outperformed even augmented versions of these algorithms.
% Two variants were tried for each method, one where the pre-trained encoders were fine-tuned and another where they were frozen. Despite these handicaps,  with the pre-trained as well as fine-tuned features (see ).
% Results are also reported for SiameseXML~\cite{Dahiya21} that uses the same text-encoder.
% Finally, results are also reported for other extreme classifiers including ECLARE~\cite{Mittal21b}, MACH~\cite{Medini2019}, Bonsai~\cite{Khandagale19}, and XT~\cite{Wydmuch18} for completeness. \\
% \todo{The final datapoint representation from these methods were used to learn a one-vs-all classifier of each label for the final relevance score. }.

\noindent\underline{\textit{A2Q-4M}}: Multi-modal baseline methods struggled to scale to this dataset so comparisons were made only to the leading text-based method SiameseXML~\cite{Dahiya21b}.

\noindent\underline {\textit{Polyvore-Disjoint}}: \alg was compared to leading methods including ADDE-O~\cite{Hou2021}, CSA-Net~\cite{lin2020}, Type-aware~\cite{Vasileva2018}, S-VAL~\cite{Kim2021}, SCE-Net average~\cite{Tan2019} and SSVR~\cite{Revanur2021}. Since this dataset offers only unseen labels as recommendation candidates at test time, only the \alg ($\alpha = 1$) variant was executed for fair comparison.

% . \alg's efficient training strategy has been shows to scale to 4 million classes while giving state-of-the-art results. \alg's gain in accuracy is also reported with respect to best baseline from extreme classification literature \ie. .

\noindent\textbf{Evaluation Metrics.}
Standard XC metrics \eg area under the curve (AUC), precision~(P@$k$), nDCG~(N@$k$), and recall~(R@$k$) were used for the MM-AmazonTitles-300K and A2Q-4M tasks. Classification accuracy was used for the multi-class Polyvore-Disjoint task as is standard~\cite{Vasileva2018, lin2020, Hou2021}.

\noindent\textbf{Hyperparameters.}
\alg uses ViT-32~\cite{dosovitskiy2021} as the image encoder $\cE_V$ with $32\times32$ patches and the msmarco-distilbert-base-v4 architecture~\cite{reimers2019} as the text encoder $\cE_T$. The AdamW optimizer with a one-cycle cosine scheduler with warm start of 1000 iterations was used. \alg could train on a 24-core Intel Skylake 2.4 GHz machine with 4 V100 GPUs within 48 hrs on the A2Q-4M dataset. See \cref{supp:hyperparameter} in the \suppciteshort for hyperparameter details.
 
\subsection{Results and Discussion}

% This demonstrates that \alg's cross attention layer in ranking stage (Phase-II) is \todo{essential} for state-of-the-art accuracy.

% does well in multiple categories - not just fashion; AttentioXML - main competitor. SiameseXML - same text encoder.

\noindent\textbf{MM-AmazonTitles-300K.} \cref{tab:mm_amz_300k} shows \alg gave 3.6--11\% higher P@1 than text-based XC methods. \alg is 5.5\% better in P@1 than AttentionXML~\cite{You18} that also employs label-specific datapoint representations. \alg is also 3.5\% more accurate than SiameseXML that uses a DistilBERT encoder similar to \alg. This indicates the benefit of melding multi-modal information with high-capacity classifier architectures. \alg's lead is similarly high in terms of other metrics such as P@5 and R@10. \alg also gave 3.2-12\% higher P@1 than multi-modal methods CLIP and VisualBERT. It is notable that the methods being compared to are variants of CLIP and VisualBERT that were offered \alg's attention modules and training strategies. \cref{supp:extra_results} shows that \alg's lead over these methods could be as high as 30\% if they are not offered these augmentations. This highlights the utility of \alg's task-specific pre-training in Module I.
% The results indicate the benefit of melding multi-modal information with high-capacity classifier architectures. \alg could be up to 30\% more accurate than leading vision + language pre-training strategies such as CLIP and VisualBERT when they are not offered fine-tuning.

% This indicates the importance of task specific pre-training in Phase-I of \alg. However, \alg dominates even when CLIP and VisualBERT are fine-tuned using \alg's training pipeline along with \ova classifiers for a fair comparison, with \alg outperforming these fine-tuned methods by 5-12\% in P@1 and 2-13\% in R@10 (see \cref{tab:mm_amz_300k}).
% indicating that \alg's cross-modal self-attention and cross-attention can lead to substantial gains over current state-of-the-art methods.

% s that \todo{using} multi-modal information in conjunction with cross-attention layer can significantly outperform leading extreme classifiers.

% Additionally, \alg uses a lighter version (\alg-light) to compare with the SiameseXML~\cite{Dahiya21b} for the fair comparison as both methods does not fine-tune datapoint embeddings and use the same text encoder. \alg-light could outperform SiameseXML by 2\% and 1.5\% in P@1 and R@10 respectively indicating, \alg is able to utilize . Similarly, \alg outperforms the best baseline by 3\% in R@$10$ by the second best method indicating superior retrieval. Please refer to table~\cref{tab:mm_amz_300k} for full details. 
% Most extreme classification methods cannot incorporate multi-modal information; for fair evaluation 

\noindent\textbf{A2Q-4M.} \alg could train on this dataset with 9M training points within 48 hrs on 4$\times$V100 GPUs. \alg achieved 47.56\% P@1 compared to 44.46\% P@1 by SiameseXML. \alg also offered predictions within 4 milliseconds per test datapoint on a single V100 GPU. \alg is able to scale to tasks with several millions of labels offering prediction times suitable for real-time applications.

\noindent\textbf{Polyvore-Disjoint.} \cref{tab:polyvore} presents results on the FITB task where \alg ($\alpha=1)$ could be 3-4\% more accurate than the next-best method. \alg is encoder-agnostic and continues to outperform existing methods even if \alg replaces its $\cE_V$ with Resnet18 (used by ADDE-O~\cite{Hou2021}).

\input{tables/table_mm_amz_3k}

\input{tables/table_polyvore_disjoint}

\begin{figure}
    \centering
    \includegraphics[width=\linewidth]{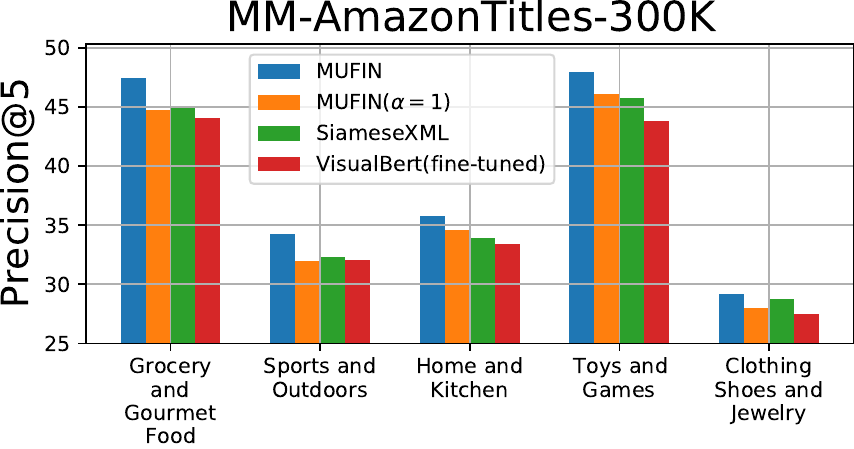}
    \caption{\alg outperforms baseline methods on almost all categories. Only the top 5 categories are shown here to avoid clutter. \cref{tab:cat_performance} in the \suppciteshort contains results on all categories.}
    \label{fig:categories}
\end{figure}

\noindent\textbf{Category-wise Analysis.} To analyze the gains offered by \alg, the performance of various algorithms was considered on the 20 unique categories of 300K products in the MM-AmazonTitles-300K dataset. \cref{fig:categories} shows that \alg's multi-modal recommendations are 2--6\%  more accurate on all popular categories than SiameseXML (that used the same text encoder $\cE_T$ as \alg). \cref{tab:cat_performance} and \cref{supp:fig:outputs} in the \suppciteshort present qualitative results that show that the trend persists and \alg offers superior performance than competing methods on almost all categories irrespective of the popularity of the category. 

% The performance of \alg was analyzed on labels from multiple categories as well as varying popularity. \todo{\alg's interpretation of inter-product cross attention and understanding of label semantics were also studied.} 

% \alg could model the intricacies of a multitude of categories.

\begin{figure}
    \centering
    \includegraphics[width=\linewidth]{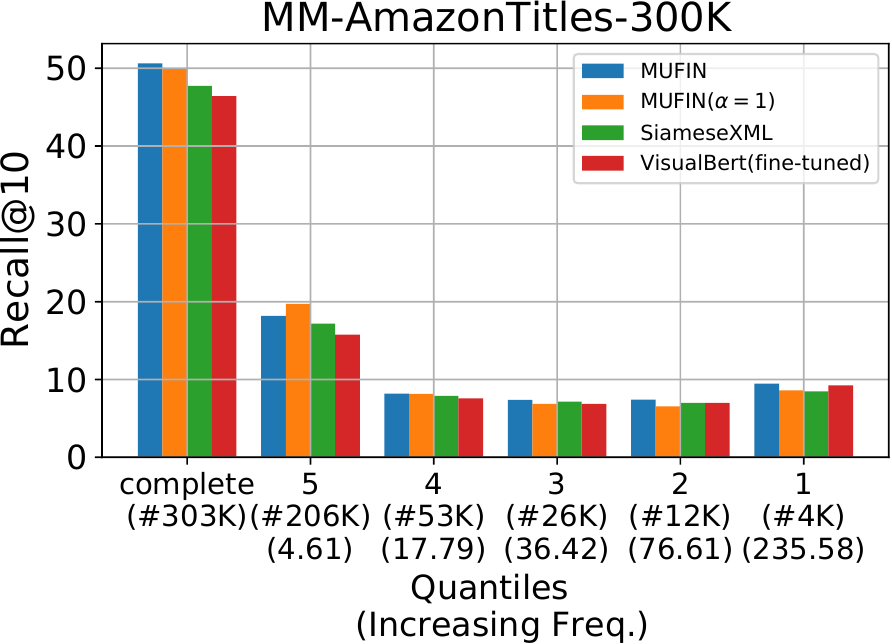}
    \caption{Analyzing the performance of \alg and other
methods on popular vs. rare labels. Labels were divided into 5 bins in increasing order of popularity (left-to-right). The plots show the overall R@10 of each method (histogram group “complete”) and the contribution of each bin to this value. The results indicate that \alg's performance on popular labels (histogram group 1) does not come at the cost of performance in rare labels. Other methods seem to exhibit a trade-off between rare and popular labels.}
    \label{fig:decile_analysis}
\end{figure}

\noindent\textbf{Label popularity.} To analyze \alg's performance on rare and popular labels, labels were divided into 5 \emph{equi-voluminous} bins of increasing label frequency such that each bin had an equal number of datapoint-label pairs from the ground truth. \cref{fig:decile_analysis} shows that \alg and \alg($\alpha=1$) outperform the baseline methods across all bins. 

\noindent\textbf{Impact of Cross Attention ($\mathcal{A}_{c}$).} \cref{fig:heat_map} shows the cross-attention heat map generated by $\cA_C$ between a datapoint $i$ and a relevant label in the retrieved shortlist $R_i$ for that datapoint. \alg was able to match a chair in a datapoint image [Image 5] to a similar chair in the background of a label image [Image 3] (magnified in \cref{fig:imgalign}). \cref{fig:xattn_vect} shows that for a given datapoint ($\widehat{\vx}^{1}$, \mycircle{blue}), cross-attention allowed \alg to generate a label-adapted datapoint representation ($\widehat{\vx}^{2,+}$, \mystar{green}) that is embedded close to the relevant label ($\vw_{+}$, \mytriangle{green}). However, the label-adapted representation of the same datapoint ($\widehat{\vx}^{2,-}$, \mystar{red}) is unmoved for an irrelevant label ($\vw_{-}$, \mytriangle{red}). Label-adapted representations allow \alg to boost the score for relevant labels and rank them higher.

\input{figs/fig_xattn}

\begin{figure}
    \centering
    \includegraphics[width=0.8\linewidth]{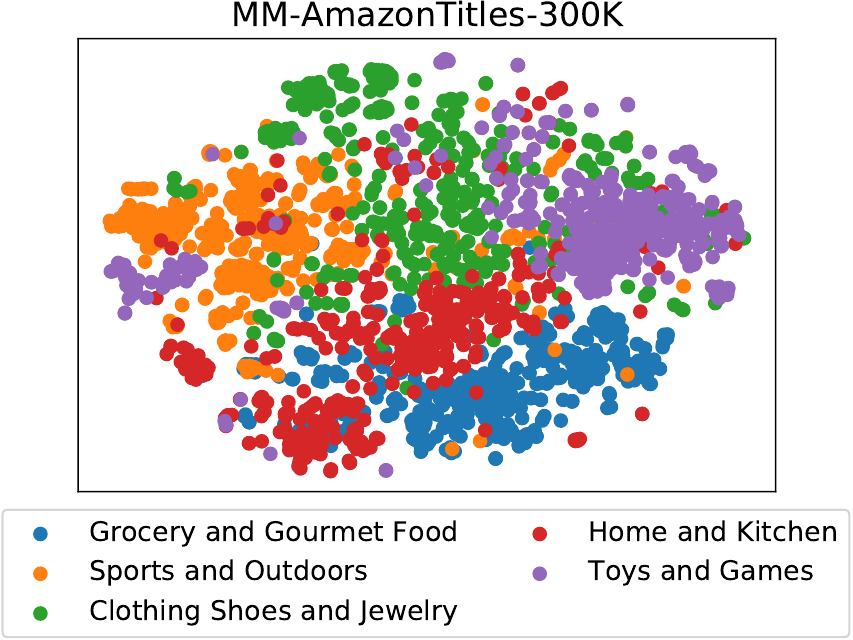}
    \caption{t-SNE representations for classifiers $\vw_l, l \in [L]$ learnt by \alg show that labels belonging to the same category are clustered together. Only the 5 most popular categories are shown.}
    \label{fig:tsne_lbl_cat}
\end{figure}

\noindent\textbf{Label semantics.} Type-aware methods~\cite{lin2020, Hou2021, Vasileva2018} have demonstrated that explicitly incorporating label categories while training can improve model accuracy. However, in XC settings with millions of labels, label hierarchies are often unavailable or incomplete~\cite{XMLRepo}. \cref{fig:tsne_lbl_cat} depicts the label classifiers ($\vw_l$) learnt by \alg using t-SNE representations. \alg could identify category-based relationships among labels without any explicit feedback on cluster identity. \alg's clusters exhibit sub-clustering that can be attributed to the fact that labels belonging to a category can be further grouped into sub-categories, \eg. \textit{``Home and Kitchen''} can be further clustered into \textit{``Furniture''} and ``\textit{Utensils}''. Thus \alg draws its gains on diverse label types (\cref{fig:categories,fig:decile_analysis}) by deducing label-datapoint relationships from multi-modal information (\cref{fig:xattn_visuals,fig:tsne_lbl_cat}).

% \alg's ability to capture such aspects of the label semantics likely \todo{helps it achieve state-of-the-art recall.}

\subsection{Ablation}
\label{sec:ablation}
This section investigates design choices made by \alg for its key components - sampling, retrieval, representation, and ranker (scorer). \cref{tab:ablation} summarizes the ablation results. The ablation experiments have been explained in detail in \cref{supp:ablation} in the \suppcite.\\
\textbf{Sampling.} Removing hard +ve sampling (\alg-no +ve) causes a 1\% drop in P@5. Removing hard -ve and +ve sampling (\alg-no +ve, -ve) leads to a 1.5\% drop in P@5.\\
\textbf{Retrieval.} Recall from \cref{sec:method} that retrieval of label shortlists $R_i$ could have been done over label embeddings $\hat\vz^1_l$ or bag embeddings $\hat\vZ^1_l$ of the labels. The augmented retrieval strategy of \alg (\alg-P-I-bag) can be 0.3\% and 1\% more accurate in R@10 and P@1 as compared to retrieval based on vector embeddings $\hat\vz^1_l$ alone (\alg-P-I-vec).\\
\noindent\textbf{Representation.} \alg was 3-4\% more accurate than the \alg-ConCat variant that concatenated $\hat\vx^1,\hat\vz^1_l$ followed by two feed-forward layers instead of using the cross-attention block $\cA_C$. Removing the self-attention block from Modules I-IV (\alg-no $\cA_S$) led to a 1.6\% drop in P@5.\\
\textbf{Ranker.} \alg's novel scoring architecture can be upto 1.5\% more accurate in terms of P@5 than variants that either exclude the cross-attention block (\alg-no $\cA_C$) or the \ova classifiers (\alg-($\alpha=1$)).

The ablations show that \alg's design choices with respect to hard +ve, -ve sampling, self- and cross-attention, and \ova classifiers, each offer performance boosts.

\input{tables/ablation}

%% file: tables/dataset_stats.tex
\begin{table*}
    \centering
    \caption{Statistics for datasets used to benchmark \alg. For Polyvore-Disjoint, a `-' indicates that only unseen labels were available for recommendation at test time. For A2Q-4M, a $\ddagger$ indicates numbers redacted for the proprietary dataset.}
    \label{tab:stats}
    \resizebox{0.9\linewidth}{!}{
    \begin{tabular}{lcccccc}
    \toprule
        Dataset & \begin{tabular}[c]{@{}c@{}}Train Datapoints \\ $N$ \end{tabular} & \begin{tabular}[c]{@{}c@{}}Labels \\ $L$ \end{tabular} & \begin{tabular}[c]{@{}c@{}}Test Instances \\ $N^{'}$ \end{tabular} & \begin{tabular}[c]{@{}c@{}}Average Labels \\ per datapoint \end{tabular} &
        \begin{tabular}[c]{@{}c@{}}Average Tokens \\ per datapoint \end{tabular} &
        \begin{tabular}[c]{@{}c@{}}Average Images \\ per datapoint \end{tabular}  \\ \toprule
        Polyvore-Disjoint & 16,995 & - & 15,145 & 1 & 27.31 & 4 \\ \hline
        MM-AmazonTitles-300K & 586,781 & 303,296 & 260,536 & 8.13 & 20.41 & 4.91\\ \hline
        A2Q-4M & 9,618,490 & 4,528,191 & 3,933,149 & $\ddagger$ & $\ddagger$ & $\ddagger$ \\\bottomrule
    \end{tabular}
    }
    
\end{table*}

%% file: tables/table_mm_amz_3k.tex
\begin{table}[!ht]
    \caption{Results on MM-AmazonTitles-300K. \alg outperforms state-of-the-art XC methods by 3--11\% in P@1 as well as R@10. \alg also outperforms state-of-the-art vision+language pre-training strategies by 3--12\% in P@1 and 4--13\% in R@10. Recall that the multi-modal techniques were augmented with \alg's pipeline. \cref{supp:extra_results} shows that \alg's lead rises significantly if the methods are not offered these augmentations. The column $t_{\text{pred}}$ reports the per-datapoint prediction times in milliseconds for various methods. \alg offers millisecond level prediction comparable to or better than existing methods.}
    \label{tab:mm_amz_300k}
    \centering
    \resizebox{\linewidth}{!}{
    \begin{tabular}{lcccccc}
    \hline
        \textbf{Method} & \begin{tabular}[c]{@{}c@{}}\textbf{P@1}/\\ \textbf{N@1} \end{tabular}  & \textbf{P@5} & \textbf{N@5} & \textbf{R@10} & \textbf{AUC}  & \begin{tabular}[c]{@{}c@{}}\textbf{$t_{\text{pred}}$}\\\textbf{(ms)} \end{tabular}  \\ \hline
        \textbf{\alg} & \textbf{52.3} & \textbf{34.76} & \textbf{50.46} & \textbf{50.63} & \textbf{0.60} & 1.32 \\ \hline
        % \alg-light & 50.88 & 33.89 & 49.14 & 49.17 & 0.59 & -- \\ \hline
        \multicolumn{7}{c}{\textbf{Textual (XC)}}\\
        \hline
        SiameseXML~\cite{Dahiya21b} & 48.64 & 32.99 & 47.46 & 47.72 & 0.57 & 0.82 \\ \hline
        ECLARE~\cite{Mittal21b} & 47.84 & 32.22 & 46.04 & 46.08 & 0.55 & 0.08\\ \hline
        AttentionXML~\cite{You18} & 46.45 & 31.17 & 44.34 & 43.73 & 0.53 & 4.33 \\ \hline
        Bonsai~\cite{Khandagale19} & 47.34 & 31.65 & 45.4 & 45.04 & 0.55 & 5.71 \\ \hline
        MACH~\cite{Medini2019} & 42.22 & 28.14 & 40.39 & 39.7 & 0.49 & 0.46 \\ \hline
        XT~\cite{Wydmuch18} & 41.45 & 27.71 & 39.64 & 38.91 & 0.52 & 5.21 \\ \hline
        \multicolumn{7}{c}{\textbf{Visual + Textual}}\\
        \hline
        \begin{tabular}[l]{@{}l@{}} CLIP~\cite{radford2021}\\+ \alg \end{tabular} & 40.49 & 27.38 & 38.45 & 37.19 & 0.485 & 6.46 \\ \hline
        \begin{tabular}[l]{@{}l@{}} VisualBert~\cite{li2019}\\+ \alg \end{tabular} & 49.11 & 32.35 & 46.95 & 46.43 & 0.58 & 8.35\\ \hline
        % CLIP~\cite{radford2021} & 22.02 & 13.83 & 20.91 & 22.27 & 0.33  \\ \hline
        % CLIP (fine-tuned) & 33.8 & 23.88 & 32.48 & 32.37 & 0.42  \\ \hline
        % VisualBert~\cite{li2019} & 2.2 & 1.4 & 2.02 & 2.4 & 0.04  \\ \hline
        % VisualBert (fine-tuned) & 42.38 & 28.61 & 41.78 & 44.44 & 0.53  \\ \hline
        
    \end{tabular}
    }
    
\end{table}

%% file: tables/table_polyvore_disjoint.tex
\begin{table}[t]
    \caption{Results on the FITB task on Polyvore-Disjoint. \alg is 3-4\% more accurate compared to the next best method.}
    \label{tab:polyvore}
    \centering
    \resizebox{0.75\linewidth}{!}{
    \begin{tabular}{lc}
    \toprule
        \textbf{Methods} & \textbf{FITB Accuracy} \\ \hline
        \textbf{MUFIN ($\alpha=1$)} & \textbf{64.17}\\ \hline
        \textbf{MUFIN (Resnet18, $\alpha=1$)} & \textbf{61.63} \\ \hline
        \multicolumn{2}{c}{\textbf{Visual + Textual}}\\ \hline
        ADDE-O~\cite{Hou2021} & 60.53 \\ \hline
        Type-aware~\cite{Vasileva2018} & 55.65 \\ \hline
        SCE-Net average~\cite{Tan2019} & 53.67\\ \hline
        SSVR~\cite{Revanur2021} & 51.5 \\ \hline
        \multicolumn{2}{c}{\textbf{Visual}}\\ \hline
        CSA-Net~\cite{lin2020} & 59.26 \\ \hline
        S-VAL~\cite{Kim2021} & 54.3 \\ \bottomrule
    \end{tabular}}
\end{table}

%% file: figs/fig_xattn.tex
\begin{figure}
    \centering
    \begin{subfigure}[t]{\linewidth}
        \vspace{0pt}
        \centering
        \includegraphics[width=0.7\textwidth]{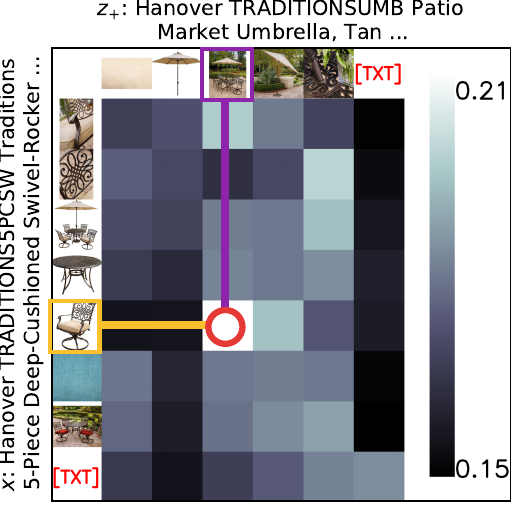}
        \caption{Heat map of cross-attention weights generated by $\cA_C$ (range of weights in legend). \textcolor{red}{{[TXT]}} denotes the title of the data-point/label. Note the point of high attention encircled in red (discussed below). The plot has been enhanced for contrast. \textbf{Please zoom in for better viewing.}}
        \label{fig:heat_map}
    \end{subfigure}
    \hfill%
    \begin{subfigure}[t]{0.54\linewidth}
        \vspace{0pt}
        \centering
        \includegraphics[width=\textwidth]{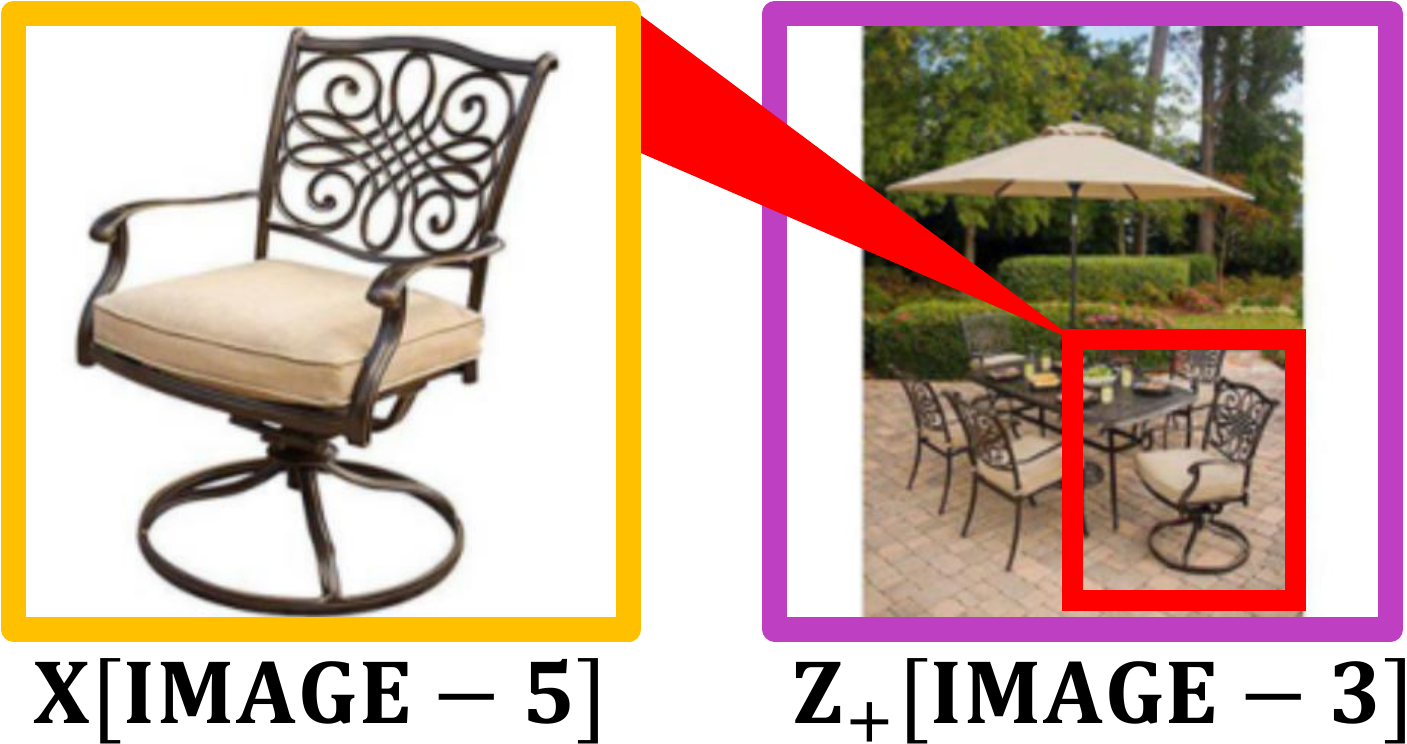}
        \caption{The high attention weight (highlighted as a red circle in \cref{fig:heat_map}) was a result of the cross-attention block $\cA_C$ being able to align [Image 5] of the datapoint $X$ to [Image 3] if the label $Z_+$ by observing that the chair depicted in [Image 5] $X$ is closely related to the chair in the background of [Image 3] of $Z_+$.}
        \label{fig:imgalign}    
    \end{subfigure}
    \hfill%
    \begin{subfigure}[t]{0.42\linewidth}
        \vspace{0pt}
        \centering
        \includegraphics[width=\textwidth]{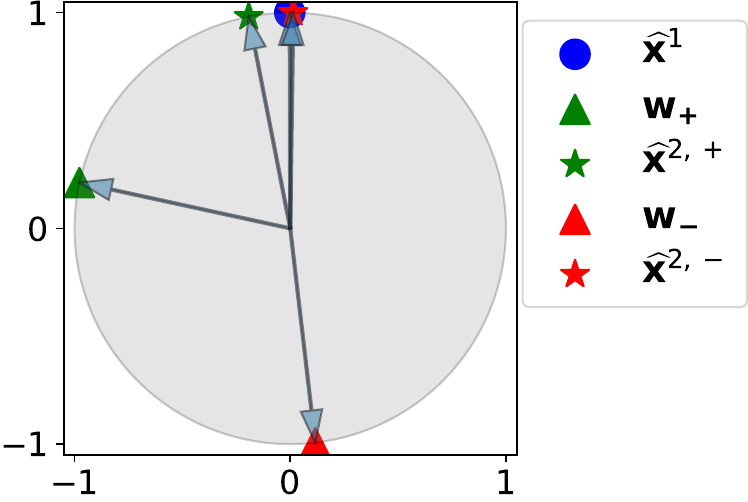}
        \caption{Analysing the impact of label-adapted representations. The vector embedding of the data point was adapted to the relevant label $Z_+$ (the very one referenced in \cref{fig:heat_map,fig:imgalign}) as well as an irrelevant label $Z_-$ (see discussion below).}
        \label{fig:xattn_vect}
    \end{subfigure}
    \caption{The impact of multi-modal cross-attention in \alg. \cref{fig:heat_map} shows the cross-attention heat map between the datapoint $X$ and a relevant label $Z_+$. \cref{fig:imgalign} shows that cross-attention is able to identify objects in $X$ among objects in the background of the relevant label $Z_+$. \cref{fig:xattn_vect} shows how this helps boost the scores assigned to relevant labels. $\hat\vx^1$ (\mycircle{blue}) represents the non-adapted vector embedding of the datapoint $X$. $\vw_+$(\mytriangle{green}) and $\vw_-$(\mytriangle{red}) represent the classifier vectors for the relevant and irrelevant labels $Z_+$ and $Z_-$ respectively. Similarly, $\hat\vx^{2,+}$(\mystar{green}) and $\hat\vx^{2,-}$(\mystar{red}) represent the vector embedding of the datapoint adapted to $Z_+$ and $Z_-$ respectively. Notice how adaptation moves $\hat\vx^{2,+}$(\mystar{green}) closer to $\vw_+$(\mytriangle{green}) allowing the relevant label $Z_+$ to get a higher score. On the other hand, adaptation has no effect when done with respect to an irrelevant label (note that $\hat\vx^1$ (\mycircle{blue}) and $\hat\vx^{2,-}$(\mystar{red}) do indeed almost overlap). \cref{fig:xattn_vect} was plotted by projecting the vectors $\hat\vx^1, \vw_+, \vw_-, \hat\vx^{2,+}, \hat\vx^{2,-}$ onto $\bR^2$ using a t-SNE embedding.}
    \label{fig:xattn_visuals}
\end{figure}

%% file: tables/ablation.tex
\begin{table}
    \centering
    \caption{An ablation study exploring alternate architecture and training choices for \alg. Choices made by \alg could lead to 3-4\% gain in P@1 and 1-2\% gain in R@10 than alternatives.}
    \resizebox{0.8\linewidth}{!}{
    \begin{tabular}{lcccc}
    \hline
        \textbf{Ablation} & \begin{tabular}[c]{@{}c@{}}\textbf{P@1}/\\ \textbf{N@1} \end{tabular} & \textbf{P@5} & \textbf{N@5} & \textbf{R@10} \\ \hline
        \textbf{\alg} & \textbf{52.3} & \textbf{34.76} & \textbf{50.46} & \textbf{50.63} \\ \hline
        \multicolumn{5}{c}{\textbf{Sampling}} \\ \hline
        \alg-no +ve & 50.35 & 33.71 & 48.91 & 49.19 \\ \hline
        \alg-no +ve, -ve & 49.69 & 33.33 & 47.9 & 48.76 \\ \hline
        \multicolumn{5}{c}{\textbf{Retrieval}} \\ \hline
        \alg-P-I-bag & 42.72 & 28.8 & 42.03 & 44.49 \\ \hline
        \alg-P-I-vec & 41.71 & 28.26 & 41.31 & 44.2 \\ \hline
        \multicolumn{5}{c}{\textbf{Representation}} \\ \hline
        \alg-ConCat & 49.61	& 32.89	& 47.87	& 47.97	\\ \hline
        \alg-no $\cA_S$ & 49.98 & 33.16 & 48.11 & 48.11 \\ \hline
        % \alg - $2\mathcal{A}_s\xrightarrow{}2\mathcal{A}_{c}$ & 50.87 & 33.93 & 49.12 & 49.11 \\ \hline
        \multicolumn{5}{c}{\textbf{Ranker}} \\ \hline
        \alg-no $\cA_C$ & 50.22 & 33.87 & 49.03 & 49.68 \\ \hline
        \alg-($\alpha=1$) & 49.25 & 33.19 & 48.53 & 49.87 \\ \hline
    \end{tabular}
    }
    
    \label{tab:ablation}
\end{table}

%% file: supp.tex
\textbf{Dataset and Supplementary Material.} The MM-AmazonTitles-300K dataset can be downloaded at \repourlcite. \alg pseudocode, implementation details, additional results and discussions on limitations of \alg, ethical considerations and future work are presented in the supplementary at \suppurlcite.

%% file: app.tex
\onecolumn
\appendix

\noindent{\large \textbf{Code and Data Release}}

\noindent Due to large size of the datafiles, the MM-AmazonTitles-300K dataset and the \alg model trained on this dataset has been released at the following URL

\suppurlcite

\noindent The \alg code has been submitted alongside this paper as supplementary material on the submission website. However, it is available on the above mentioned URL as well.\\

\noindent{\large \textbf{Discussion on Ethical Considerations, Limitations and Future Work}}

\noindent\textbf{Ethical Considerations}: The MM-AmazonTitles-300K dataset was curated from a publicly available crawl~\cite{ni2019} and utilizes no personally identifiable or human subject data. The \alg approach is itself applied to tasks such as product-to-product recommendation, bid query prediction, and outfit completion that seek to improve user experience when browsing for products. We are unaware of any direct applications of the \alg approach that can have negative societal impact.

\noindent\textbf{Limitations and Future Work}: We identify two potential avenues for further improving \alg's performance. Firstly, \cref{fig:decile_analysis} indicates that the \alg-$(\alpha = 1)$ is better at predicting tail/rare labels correctly whereas \alg performs better on head/popular labels. Although \alg outperforms \alg-$(\alpha = 1)$ overall (see \cref{tab:ablation}), this indicates towards a possibility for a third variant that scores rare and popular labels differently to get the best of both worlds. Secondly, it is common for products on e-commerce and other portals to be endowed with taxonomies that can give direct information about related products. The use of such tree/graph metadata has been shown in XC literature \cite{Mittal21b} to positively impact performance. Incorporating such relational metadata at the scale of millions of products is an interesting direction.

\noindent

\section{\alg Pseudocode and Outline}
\label{supp:pseudocode}
\cref{fig:pipeline} below presents \alg's prediction pipeline. The ViT+SentenceBERT encoder $\cE$ trained in a Siamese fashion embeds a test point $X$ as a vector $\hat\vx^1$ which is used to shortlist $O(\log{L})$ labels using augmented retrieval. For each shortlisted label e.g. label number 42 in the example, a label-adapted representation $\hat\vx^{2,42}$ is generated by applying cross-attention between test point representation $\hat\vX^1$ and label representation $\hat\vZ_{42}^1$. The dot product between $\hat\vx^{2,42}$ and the classifier for this label $\vw_{42}$ gives final score of label 42 for this test point.

\begin{figure}[h]
    \centering
    \includegraphics[width=\linewidth]{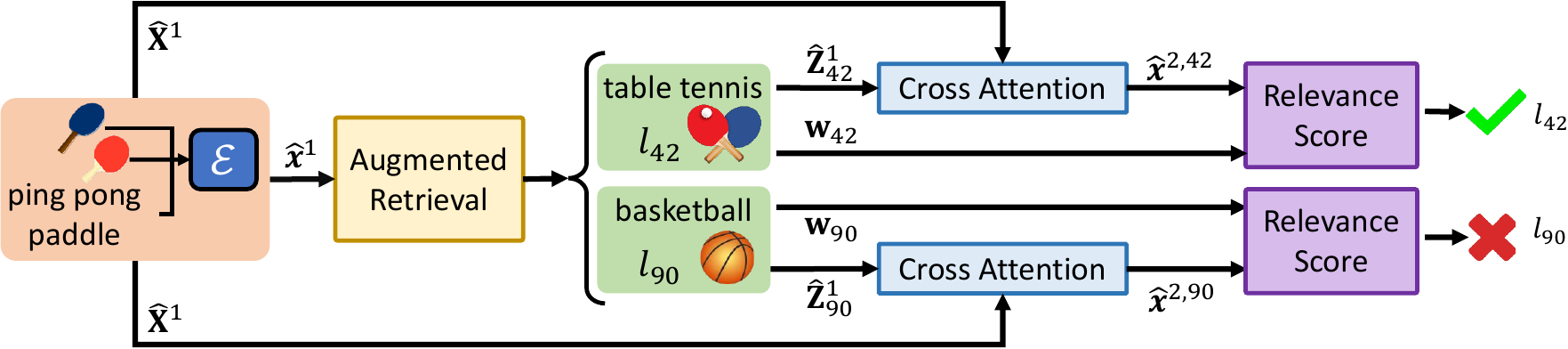}
    \caption{Scalable prediction pipeline deployed by \alg}
    \label{fig:pipeline}
\end{figure}

\noindent\textbf{Prediction Time Complexity for \alg.} For sake of simplicity, let us assume that all labels and datapoints contain the same number $m$ of descriptors. Computing the bag of pre-embeddings $\hat\vX^0_t$ takes $\bigO{m\cdot(\text{ENC} + D)}$ time assuming the encoders $\cE_V,\cE_T$ take $\text{ENC}$ time to encode a descriptor into a $D$-dimensional vector. Passing $\hat\vX^0_t$ through the self-attention block $\cA_S$ to obtain $\hat\vX^1_t$ takes $\bigO{mD^2 + m^2D}$ time after which computing the vector representation $\hat\vx^1_t$ takes an additional $\bigO{mD}$ time. Querying the ANNS structure $\anns$ to obtain the shortlist $R_t$ of $\bigO{\log L}$ labels takes at most $\bigO{D\log L}$ time \cite{MalkovY16}. Applying the cross-attention block $\cA_C$ with respect to each of these shortlisted labels takes a total of $\bigO{(mD^2 + m^2D)\log L}$ time. Vectorizing these outputs to obtain $\hat\vx_t^{2,l}$ and applying the label classifiers $\vw_l$ takes at most $\bigO{mD\log L}$ time. This brings the total time complexity of the prediction pipeline per test datapoint to be at most $\bigO{m\cdot\text{ENC} + (m + D)\cdot mD\log L}$. In practice, \alg offered predictions within a 3-4 milliseconds even on tasks such as A2Q-4M with several millions of labels.

\section{Datasets}
\label{supp:datasets}

\noindent(Discussion continued from the \textbf{Datasets} subsection in \cref{main:dataset}) \\

\noindent\underline {\textit{MM-AmazonTitles-300K.}}: This dataset was curated from an existing Amazon click dump~\cite{ni2019}. Given a product as a data-point, the aim is to recommend the subset of the most relevant (e.g. frequently bought-together) products from a catalog of over 300K unique products. Each product is represented using multiple descriptors including a product title and up to 15 product images. The dataset released in the \suppcite consists of multiple product entries in JSON format. Each product is associated with multiple tags described below:
\begin{enumerate}
    \item ``ASIN'': this acts as a unique identifier (UID) for the product
    \item ``title'': this represents a textual title for the product
    \item ``images'': this presents a list of URLs pointing to multiple (up to 15) images of the product
    \item ``also\_buy'': this presents a list of UIDs of products that were frequently bought together with the product
\end{enumerate}
Products having no images as well as no title were not included in the dataset. To generate the train test split, guidelines from~\cite{XMLRepo} were closely followed.

\section{Baselines and Related work}
\label{supp:baseline}

\noindent(Discussion continued from the  \textbf{Baselines} subsection in \cref{main:baselines})\\

\noindent\textbf{Baselines for the Polyvore-Disjoint dataset}: Performance numbers for all baselines in \cref{tab:polyvore} were taken directly from published results~\cite{Hou2021, Kim2021, Revanur2021, lin2020}. For sake of completion, each baseline method for this dataset is described in below in brief.
\begin{itemize}
\item\textbf{ADDE-O~\cite{Hou2021}}: Learns attribute-driven disentangled representations.

\item\textbf{CSA-NET~\cite{lin2020}}: Learns a category-based subspace attention network for scalable indexing and retrieval. This work introduced the notion of \textit{outfit ranking loss} that considers the item-relationship of an entire outfit.

\item\textbf{Type-aware~\cite{Vasileva2018}}: Learns an image embedding that respects item type, and jointly learns notions of item similarity and compatibility in an end-to-end manner. This method uses both visual and textual descriptors to represent an outfit.

\item\textbf{S-VAL~\cite{Kim2021}}: Learns outfit representation by self-supervision. The self-supervision tasks used in the paper were histogram prediction and learning to discriminate shapeless patches and textures from different images.

\item\textbf{SCE-Net~\cite{Tan2019}}: Learns model parameters by optimizing a combination of loss functions. For Polyvore-Disjoint, SCE-Net uses two objective loss function, namely VSE and Sim. The VSE loss requires that image embeddings and textual embeddings of the same outfit be embedded together. The Sim loss encourages images and descriptions of similar products to be embedded close to each other.

\item\textbf{SSVR-Net~\cite{Revanur2021}}: Learns model parameters using semi-supervised learning. This work uses a Siamese architecture trained using triplet loss that takes an input image, a positive instance (an affine transformation of the image) and a negative instance (an image with color transformations such as random gray scaling, jittering \etc).
\end{itemize}

\noindent\textbf{Baselines for the MM-AmazonTitles-300K dataset}: Baselines for this dataset are divided in two sets.
\begin{enumerate}
    \item \textbf{Textual Methods}: Methods in the first set correspond to state-of-the-art extreme classification techniques including, AttentionXML, SiameseXML, ECLARE, Bonsai, MACH and XT. These methods are designed to be text-based and as such use only textual descriptors of a product. Consequently, each product was represented for these methods using its product title alone. Hyperparameters for each method were used as suggested by the respective papers. With the exception of the MACH method, all these methods enjoy an $\bigO{\log L}$ prediction time complexity similar to \alg.
    \item \textbf{Visual + Textual Methods}: These methods include CLIP~\cite{radford2021} and VisualBert~\cite{li2019} that use both visual and textual descriptors of a product.
\end{enumerate}
As before, each baseline method for this dataset is described in below in brief. The details of how CLIP and VisualBERT were augmented and fine-tuned are discussed thereafter.

\begin{itemize}

\item\textbf{SiameseXML~\cite{Dahiya21b}}: Melds Siamese networks with \ova classifiers. SiameseXML retrieves label shortlists using multiple ANNS structures unlike \alg that uses a single ANNS structure. The shortlisted labels are then ranked according to scores obtained from label-wise \ova classifiers.

\item\textbf{ECLARE~\cite{Mittal21b}}: Exploits label graphs to obtain superior label representations with an aim to improve performance on rare labels. Since the MM-AmazonTitles-300K dataset does not provide a label graph natively, a label graph was mined by performing random walks using the label vectors as suggested in \cite{Mittal21b}.

\item\textbf{AttentionXML~\cite{You18}}: Learns to partition labels using a shallow and wide PLT (depth between 2-3). A context vector is learnt per label that is used to generate label-specific datapoint representations. We note that the cross-attention block $\cA_C$ used by \alg similarly produces label-adapted datapoint representations.

% Explores label co-occurrence graph to improve predictions on rare labels. ECLARE uses graph augment label centroids to partition the label space. At the time of retrieval, the shortlist of labels is expanded by including labels which co-occur in the graph. During learning model parameters, labels which co-occur, learn \ova classifier by sharing gradients of each other.

\item\textbf{Bonsai~\cite{Khandagale19}}: Implements a scalable tree-based classifier by learning a label hierarchy over the labels by representing each label using its bag-of-words (BoW) centroid vectors.

\item\textbf{MACH~\cite{Medini2019}}: Learns an ensemble of 32 learners where each learner randomly partitions labels into several hash bins. Models are learnt to predict the hash bits corresponding to each learner. At prediction time, a majority vote is taken over all the learners to boost confidence. The method offers a prediction time of $\bigO L$.

\item\textbf{XT~\cite{Wydmuch18}}: Generalizes the hierarchical softmax approach popular for multi-class problems to multi-label problems using a probabilistic label tree. Recall from the discussion in \cref{sec:intro} that in multi-class classification, the objective is to predict a single mutually exclusive label for a given datapoint whereas in multi-label classification, the goal is to annotate datapoints with the most relevant \textit{subset} (one or more) of labels.

\item\textbf{VisualBert~\cite{li2019}}: Uses pre-trained Resnet-101 embeddings to represent images. Given a data point with multiple visual and textual descriptors, the visual descriptors are encoded and fed as tokens into a BERT (large) architecture alongside textual tokens from the textual descriptors. In this sense, VisualBert can be seen as utilizing early fusion. 

\item\textbf{CLIP~\cite{radford2021}}: Uses a ViT architecture to encode images and a BERT architecture to encode text. The method pre-trains the architectures to encode relevant image-text pairs together. Subsequently, a late fusion architecture is learnt that fuses an image embedding and a text embedding into a joint representation over which a scoring model can be learnt.

\end{itemize}

\newcommand{\CP}{{\text{CP}}}
\newcommand{\VB}{{\text{VB}}}

\subsection{Adapting CLIP and VisualBERT to MM-AmazonTitles-300K and Fine-tuning Details}

\noindent\textbf{Datapoint/label embedding Architecture}: As mentioned above, CLIP uses separate encoders to encode images and text that can offer a bag of embeddings $\cE_\CP(X_i) \in \bR^{m_i \times D}$. This bag of embeddings was fed into a fresh instantiation of the self-attention block $\cA_{S}$ to obtain datapoint/label embeddings that we name $\hat\vx^\CP$ and $\hat\vz^\CP_l$. These are analogous to the $\hat\vx^1$ and $\hat\vz^1_l$ representations used by \alg. On the other hand, VisualBert uses a BERT architecture and pre-trained Resnet101 embeddings for images as tokens along with textual tokens to itself encode each datapoint/label as a vector to give embeddings $\hat\vx^\VB$ and $\hat\vz^\VB_l$. Note that VisualBERT was not offered the self-attention block since it itself performs similar self-attention operations within the layers of its BERT (large) architecture.

\noindent\textbf{Retrieval}: For VisualBERT, retrieval was performed using an ANNS structure created over the label embeddings $\hat\vz^\VB_l$. For CLIP, since it offers bag embeddings $\cE_\CP(Z_L) \in \bR^{m_l \times D}$ one per descriptor of the label, augmented retrieval was implemented similar to \alg.

\noindent\textbf{Scoring Architecture}: Both CLIP and VisualBERT were offered independent instantiations of the cross-attention block $\cA_C$. For CLIP, it was applied to the bag embeddings $\cE_\CP(X_i)$ and $\cE_\CP(Z_l)$. For VisualBERT that does not offer bag embeddings, cross-attention was applied over $\hat\vx^\VB_i$ and $\hat\vz^\VB_l$ instead. Both methods were offered \ova classifiers $\vw_l, l \in [L]$.

\noindent\textbf{Training and Fine-Tuning Details}: In the first experiment, the encoding architectures of CLIP and VisualBERT were frozen and only the self-/cross-attention architectures were trained in a four module manner identical to \alg. The results of this experiment correspond to the rows titled ``VisualBert'' and ``CLIP'' in \cref{tab:mm_amz_300k}. In the next experiment, the encoders used within CLIP and VisualBERT were additionally trained in Modules I-IV. The results of this experiment correspond to the rows titled ``VisualBert-fine-tuned'' and ``CLIP-fine-tuned'' in \cref{tab:mm_amz_300k}. Without fine-tuning the encoders, both architectures offered poor performance 30-50\% worse than \alg in terms of P@1. Fine-tuning significantly improved the performance for both methods but they remained 5-13\% worse than \alg in terms of P@1.

% In Module I the base models (CLIP and VisualBert) were fine-tuned using the respective representations similar to \alg. At the retrieval stage CLIP used \alg's bag retrieval while VisualBert could not use the bag retrieval as it generates a single datapoint representation. During Module IV training, CLIP and VisualBert were given \alg's cross-attention based extreme classifiers for final ranking. CLIP and VisualBert were given an additive advantage of self attention and cross attention whenever architecture of the base model permits. Fine-tunting using \alg's training pipeline lead to a significant gain is accuracy of 25-40\% over the base models. However, \alg could outperform, CLIP-fine-tuned and VisualBert-fine-tuned by 5-12\% in P@1.

\section{Hyperparameters}
\label{supp:hyperparameter}
\input{tables/hyper_parameters}

\noindent(Discussion continued from the \textbf{Hyperparameters} subsection in \cref{main:dataset})\\

\noindent \alg uses a ViT-32~\cite{dosovitskiy2021} architecture as the image encoder $\cE_V$ with $16\times16$ patches and a SentenceBert~\cite{reimers2019} architecture as the text encoder $\cE_T$. The AdamW optimizer with a one-cycle cosine scheduler with warm start of 1000 iterations was used. \alg was trained on a 24-core Intel Skylake 2.4 GHz machine with 4 V100 GPUs for 48 hrs on the A2Q-4M dataset. To reiterate the main training parameters mentioned in \cref{sec:method}, model parameters were learnt in Module I using the contrastive loss with a margin of $\gamma = 0.2$. In Module IV, these encoders were fine-tuned using the cosine embedding loss with a margin of $\gamma = 0.5$ with $\abs{\cS_i} = 2$ randomly sampled positive labels and $\abs{\cT_i} = 12$ hard negative labels taken from the shortlist $R_i$ obtained from the augmented retrieval step. For details of other hyper-parameters please refer to \cref{tab:hyperparameter}.

\section{Additional Experimental Results}
\label{supp:extra_results}
\cref{supp:fig:outputs} presents predictions by \alg for a few sample test points in the MM-AmazonTitles-300K dataset. Not only does \alg mostly recommend products relevant according to the ground truth, but the occasional recommendation not in the ground truth list is often a good recommendation nevertheless.

\cref{tab:cat_performance} presents \alg's performance on various categories of the MM-AmaonTitles-300K dataset of which a snapshot was provided in \cref{fig:categories}. For most categories, \alg is indeed the best method and \alg $(\alpha = 1)$ is the second-best method. In particular, \alg could give accuracy gains up to 6\% on various categories (e.g. Musical Instruments).

\input{figs/img}

\input{tables/category_prec_5}

\section{Ablation}
\label{supp:ablation}
\noindent(Discussion continued from \textbf{Ablation} \cref{sec:ablation})\\

\alg makes several design choices with respect to key components in its architecture and training pipeline.  its key component- sampling, retrieval, representation and ranker. These ablation experiments attempt to ascertain the differential contribution of each of these design choices to \alg's final performance. The results of the ablation experiments are detailed in \cref{tab:ablation}.

\noindent\textbf{Sampling}: Recall that in Module I, \alg uses hard negative as well as hard positive sampling to focus training on datapoint-label pairs that offer the most prominent gradients. In the \alg-no +ve variant, hard positive sampling was replaced with random positives. In the \alg-no +ve, -ve variant, both hard positive and hard negative sampling was replaced with random positive and negatives. As \cref{tab:ablation} indicates, ``\alg-no +ve'' and ``\alg-no +ve, -ve'' lead to the loss of 2\% and 2.5\% in P@1 and 1\% and 1.5\% in P@5. This points to the need for effective training using hard negatives and positives.

\noindent\textbf{Retrieval}: Recall that to perform augmented retrieval, \alg represents each label using its corresponding bag-of-embeddings of product images as well as text and creating ANNS structures over an expanded set of $\sum_{l\in[L]}m_l$ vectors, the label $l$ getting represented using $m_l$ vectors, one per descriptor. We refer to this default variant as \alg-P-I-bag. The alternate variant \alg-P-I-vec represents each label using a single embedding namely $\vz^1_l$. Results shows that \alg-P-I-bag could be 1\% and 1.6\% more accurate in P@1 and P@5 indicating the benefits of augmented retrieval.\\
\textit{\textbf{Note about the Retrieval ablation experiment}}: The performance numbers for \alg-P-I-bag and \alg-P-I-vec given in \cref{tab:ablation} are those of the \alg model learnt after Module I i.e. sans the cross-attention layer and \ova classifiers. This is why the absolute numbers for \alg-P-I-bag (e.g. 42.72 P@1) are lower than \alg (52.3 P@1). The difference can be attributed to the inclusion of the cross-attention block and \ova classifiers.

% At the time of retrieval for datapoint/label pair with the high datapoint/bag-item similarity is ranked higher by \alg (). \alg experimented with the traditional retrieval scenario in which labels were represented by the embedding-of-bag ($\vz^{1}_{l}$) instead (). 

\noindent\textbf{Representation}: \alg adapts the representation of a datapoint to a label using its cross attention block $\cA_C$. In the \alg-ConCat variant, the cross attention block was replaced with a simpler architecture that concatenates the datapoint ($\vx^1$) and label ($\vz^1_l$) representations and applies two feedforward layers of the form $2D \rightarrow 2D \rightarrow D$ to obtain an alternate label-adapted representation analogous to $\hat\vx^{2,l}$. Results show that \alg could be 2.69\% and 1.87\% more accurate than \alg-ConCat in terms of P@1 and P@5. The \alg-no $\cA_S$ variant replaces the self-attention block $\cA_S$ with a simple feed-forward layer (in Modules I-IV) and observed a loss of 2.3\% and 1.6\% in terms of P@1 and P@5. The ablations indicate the effectiveness of \alg's self- and cross-attention compared to alternatives.

\noindent\textbf{Ranker}: \alg uses a cross-attention layer and \ova classifiers to perform datapoint-label scoring. In the \alg-no $\cA_C$ variant, the cross-attention block is completely bypassed, effectively yielding $\hat\vx^{2,l} = \hat\vx^1$ on top of which \ova classifiers are then applied. In the \alg-(\alg=($\alpha=1$)) on the other hand, \alg completely ignores the \ova classifiers by explicitly setting $\alpha = 1$. Experiments show that \alg scoring mechanism employing both a cross-attention block and \ova classifiers can be 2-3\% more accurate in P@1 than the \alg-no$\cA_{C}$ and \alg-($\alpha = 1$) variants.

%% file: tables/hyper_parameters.tex
\begin{table}
    \centering
    \caption{Hyper-parameters used to train \alg on the public datasets. \alg uses the AdamW optimizer and learning rate scheduler with a warm start of 1000 iterations.}
    \resizebox{0.7\linewidth}{!}{
    \begin{tabular}{|l|c|c|c|c|c|c|}
    \hline
        \multirow{2}*{Dataset} & \multicolumn{3}{c|}{Module I} & \multicolumn{3}{c|}{Module IV} \\ \cline{2-7}
         & Epochs & \begin{tabular}[c]{@{}c@{}} Learning\\ Rate (lr) \end{tabular} & \begin{tabular}[c]{@{}c@{}} Batch\\ Size (B) \end{tabular} & Epochs & \begin{tabular}[c]{@{}c@{}} Learning\\ Rate (lr) \end{tabular} & \begin{tabular}[c]{@{}c@{}} Batch\\ Size (B) \end{tabular} \\ \hline
        MM-AmazonTitles-300K & 200 & 2e-4 & 1024 & 20 & 5e-5 & 200 \\ \hline
        Polyvore-Disjoint & 200 & 2e-4 & 1024 & 10 & 5e-5 & 200 \\ \hline
    \end{tabular}
    }
    
    \label{tab:hyperparameter}
\end{table}

%% file: figs/img.tex
\begin{figure}
    \begin{subfigure}{0.18\linewidth}
        \includegraphics[height=0.75\textwidth]{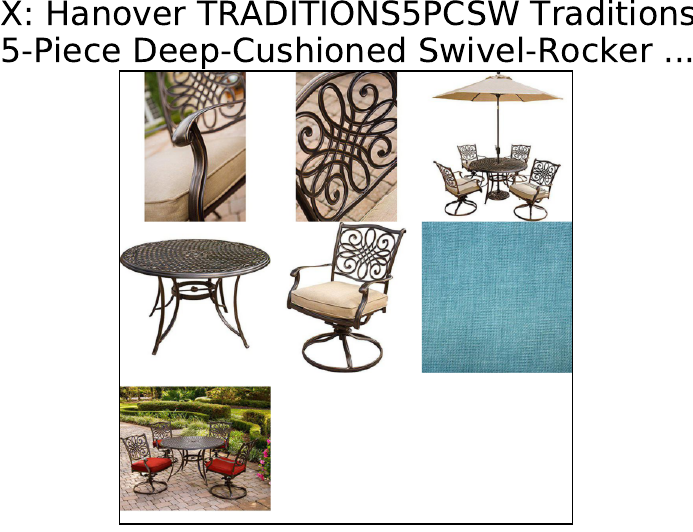}
        \newline
        \newline
        \includegraphics[height=0.75\textwidth]{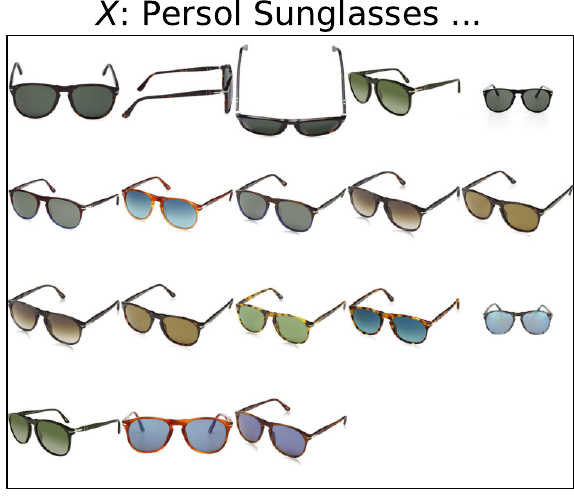}
    \caption{Query Datapoint}
    \end{subfigure}
    \hfill%
    \begin{subfigure}{0.75\linewidth}
        \includegraphics[height=0.18\textwidth]{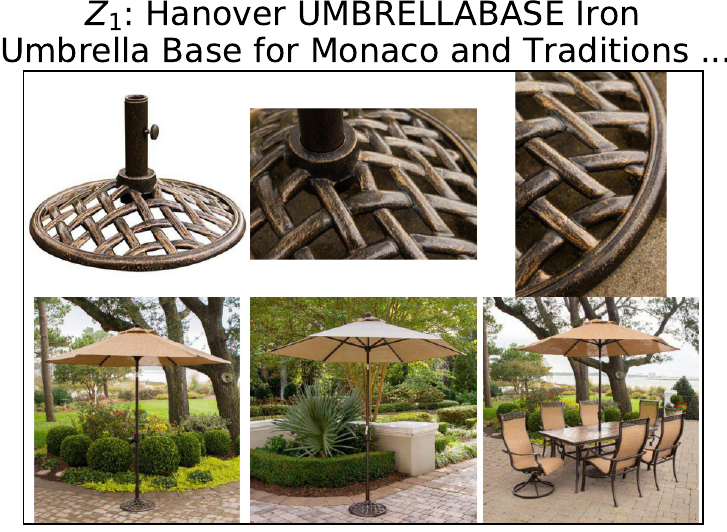}
        \includegraphics[height=0.18\textwidth]{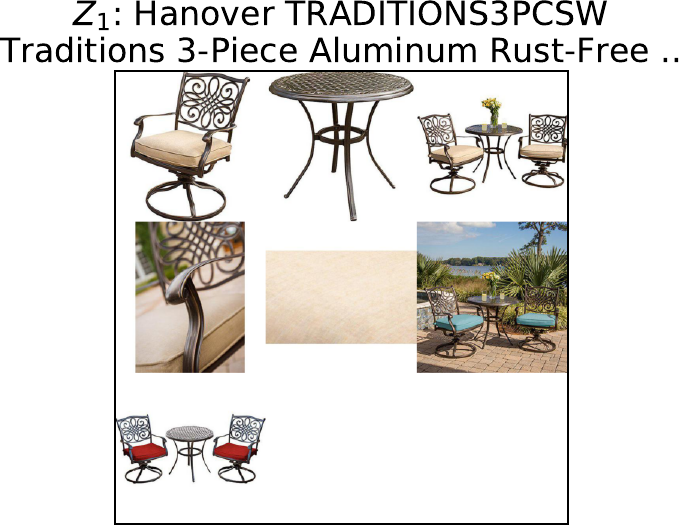}
        \includegraphics[height=0.18\textwidth]{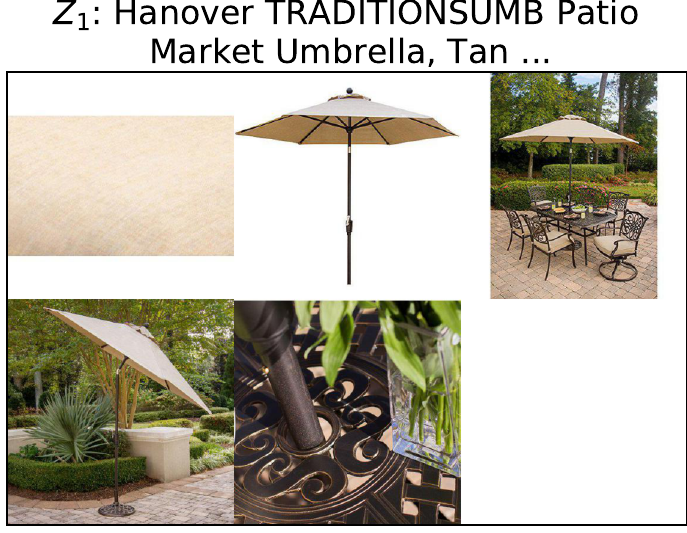}
        \includegraphics[height=0.18\textwidth]{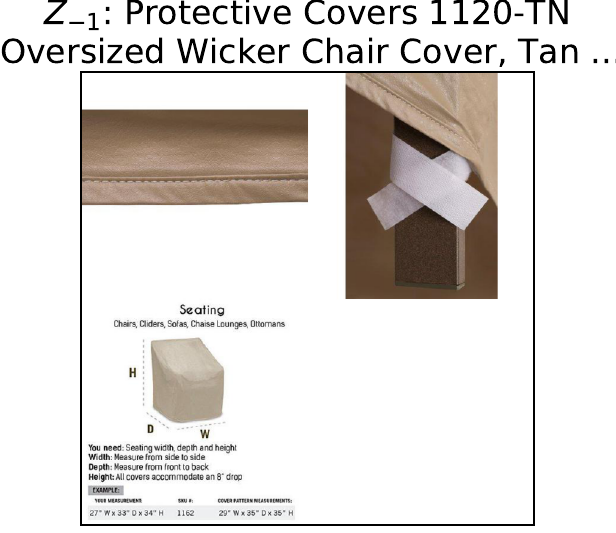}
        \newline
        \newline
        \includegraphics[height=0.18\textwidth]{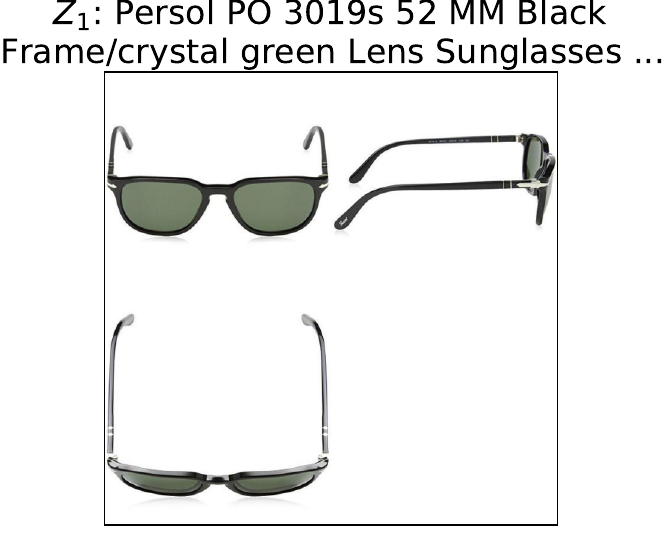}
        \includegraphics[height=0.18\textwidth]{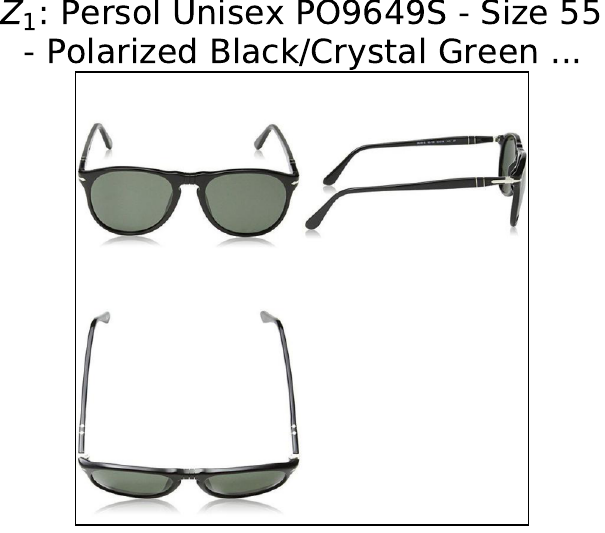}
        \includegraphics[height=0.18\textwidth]{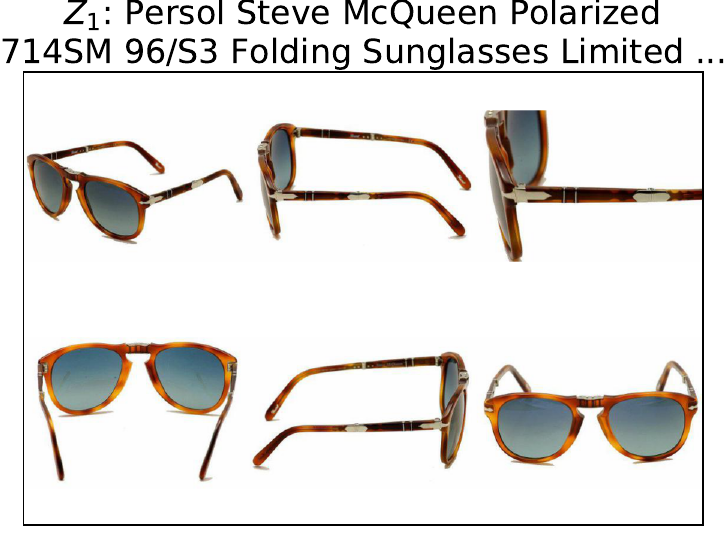}
        \includegraphics[height=0.18\textwidth]{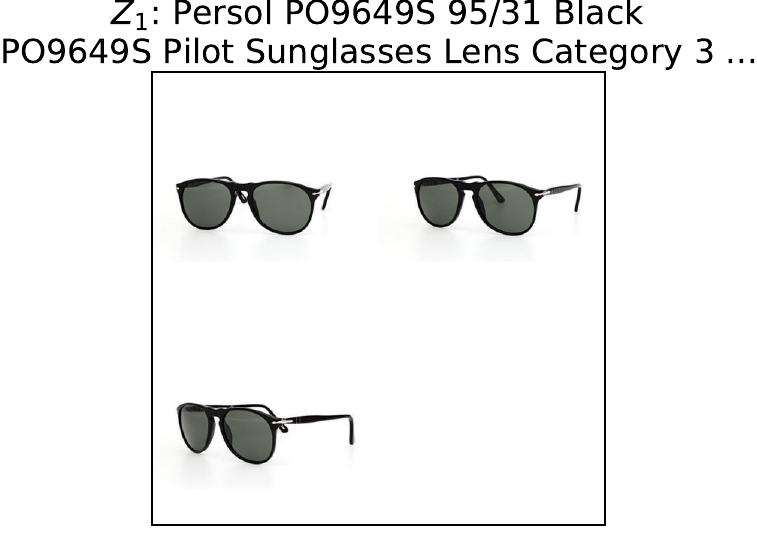}
        \caption{Retrieved results}
    \end{subfigure}
    
    \caption{Predictions by \alg for sample test points in the MM-AmazonTitles-300K dataset. $\mathbf{Z}_1$  (resp. $\mathbf{Z}_{-1}$) in the title indicates that the retrieved product (label) was relevant (resp. irrelevant) for the query datapoint in the ground truth. Not only does \alg mostly recommend products relevant according to the ground truth, but the occasional recommendation not in the ground truth list is often a good recommendation nevertheless. For instance, in the first row, the last recommendation does not appear in the ground truth but is a chair cover relevant to the query. (Figure best viewed under magnification)}
    \label{supp:fig:outputs}
\end{figure}

%% file: tables/category_prec_5.tex
\begin{table}[t]
    \centering
    \caption{\alg's performance on various categories of the MM-AmaonTitles-300K dataset of which a snapshot was provided in \cref{fig:categories}. For each category, the best performance is highlighted in bold black font, the second-best performance is left in normal black font and the third-/fourth- performances are stylized in light gray. Note that for most categories, \alg is indeed the best method and \alg $(\alpha = 1)$ is the second-best method. In particular, \alg could give accuracy gains up to 6\% on various categories (e.g. Musical Instruments).}
    \label{tab:cat_performance}
    \resizebox{0.9\linewidth}{!}{
    \begin{tabular}{|l|c|c|c|c|}
    \hline
        ~ & \# Labels & \begin{tabular}[c]{@{}c@{}} MUFIN /\\ MUFIN ($\alpha=1$) \end{tabular} & SiameseXML~\cite{Dahiya21b} & VisualBert~\cite{li2019} \\ \hline
        Overall & 303,296 & \textbf{34.76} / 33.19 & \wpred{32.99} & \wpred{32.24} \\ \hline
        AMAZON FASHION & 681 & \textbf{24.71} / \wpred{23.85} & 23.90 & \wpred{22.98} \\ \hline
        All Beauty & 329 & \textbf{19.70} / 19.04 & \wpred{18.98} & \wpred{18.80} \\ \hline
        Appliances & 767 & \textbf{27.51} / 27.17 & \wpred{24.60} & \wpred{25.69} \\ \hline
        Arts Crafts and Sewing & 16,843 & \textbf{41.62} / 39.18 & \wpred{39.06} & \wpred{38.25} \\ \hline
        Automotive & 18,384 & \textbf{14.39} / 14.12 & \wpred{12.84} & \wpred{13.40} \\ \hline
        Cell Phones and Accessories & 6,410 & \textbf{33.32} / 32.81 & \wpred{32.52} & \wpred{32.06} \\ \hline
        Clothing Shoes and Jewelry & 25,379 & \textbf{29.18} / \wpred{28.03} & 28.75 & \wpred{26.80} \\ \hline
        Electronics & 22,449 & \textbf{32.65} / 31.07 & \wpred{29.92} & \wpred{29.96} \\ \hline
        Grocery and Gourmet Food & 24,676 & \textbf{47.41} / \wpred{44.72} & 44.94 & \wpred{44.18} \\ \hline
        Home and Kitchen & 37,111 & \textbf{35.75} / 34.63 & \wpred{33.88} & \wpred{34.19} \\ \hline
        Industrial and Scientific & 7,333 & \textbf{36.55} / 34.01 & \wpred{33.77} & \wpred{33.19} \\ \hline
        Luxury Beauty & 3,455 & 60.74 / \wpred{58.98} & \textbf{61.56} & \wpred{60.72} \\ \hline
        Musical Instruments & 3,835 & \textbf{26.18} / 24.49 & \wpred{20.82} & \wpred{23.74} \\ \hline
        Office Products & 16,763 & \textbf{38.09} / 36.33 & \wpred{36.13} & \wpred{35.17} \\ \hline
        Patio Lawn and Garden & 11,631 & \textbf{36.33} / 35.03 & \wpred{32.84} & \wpred{34.25} \\ \hline
        Pet Supplies & 12,102 & \textbf{41.76} / 40.03 & \wpred{37.54} & \wpred{38.47} \\ \hline
        Prime Pantry & 3,328 & \wpred{37.45} / 38.78 & \textbf{40.21} & \wpred{38.61} \\ \hline
        Sports and Outdoors & 24,842 & \textbf{34.24} / \wpred{31.99} & 32.30 & \wpred{30.62} \\ \hline
        Tools and Home Improvement & 23,437 & \textbf{37.03} / 35.03 & \wpred{34.87} & \wpred{34.11} \\ \hline
        Toys and Games & 43,541 & \textbf{47.92} / 46.13 & \wpred{45.73} & \wpred{45.55} \\ \hline
    \end{tabular}
    }
    
\end{table}